# Photoacoustic phase-controlled Fourier-transform infrared spectroscopy

SANTERI LARNIMAA,[1,*] MIKHAIL ROIZ,[1] AND MARKKU VAINIO[1,2]

[1]*Department of Chemistry, University of Helsinki, P.O. Box 55, FI-00014 Helsinki, Finland*
[2]*Photonics Laboratory, Physics Unit, Tampere University, FI-33014 Tampere, Finland*
*\*santeri.larnimaa@helsinki.fi*

**Abstract:** We demonstrate a 13-fold speed improvement in broadband cantilever-enhanced photoacoustic spectroscopy (CEPAS) by combining it with phase-controlled Fourier-transform spectroscopy (PC-FTS) instead of traditional Fourier-transform infrared spectroscopy (FTIR). PC-FTS is a modification of FTIR and capable of fundamentally faster interferogram acquisitions. The speed-improvement is beneficial for CEPAS, which is an especially sensitive version of the background-free photoacoustic spectroscopy technique. We used the PC-FTS-CEPAS technique to measure the absorption spectrum of methane in the mid-infrared region (3.3–3.5 μm) with an optical frequency comb as the light source.

## 1. Introduction

Fourier-transform infrared spectroscopy (FTIR) is one of the most widely used spectroscopic techniques. It is based on the Michelson interferometer, where the input light is split into two arms, back reflected, and combined again [1,2]. The optical field in one of the arms is delayed by scanning the arm length with a translating mirror. This leads to either constructive or destructive interference between the combined beams, depending on the optical path difference, and maps each optical frequency into an easily measurable and unique acoustic or radio frequency. The resulting time domain signal (the interferogram) is Fourier-transformed to yield the spectrum.

Using rotational instead of translational motion in the delay arm can lead to significantly higher interferogram acquisition rates [3,4]. An especially interesting new technique based on rotational motion is phase-controlled Fourier-transform spectroscopy (PC-FTS) first presented by Hashimoto et al. [5,6]. In PC-FTS, the delay arm contains a grating that disperses the input light onto a rotating mirror surface enhancing the delay obtainable with the tilt of the mirror. One can choose which optical frequency is down converted to the zero radio frequency, which leads to efficient use of the detection bandwidth and fundamentally faster interferogram acquisition rates than traditional FTIR. The PC-FTS method has been successfully used in the near-infrared and in the mid-infrared regions for gas-phase spectroscopy and to monitor dynamic mixing of liquids with interferogram acquisition rates as high as 24 kHz [5,6].

We take a different approach and utilize the speed benefit of PC-FTS in cantilever-enhanced photoacoustic spectroscopy (CEPAS). Photoacoustic spectroscopy (PAS) is a highly sensitive spectroscopic technique that is widely used in the analysis of condensed matter [7–12], aerosols [13–15], and molecular gases [16–19]. Here, we mainly focus on trace gas analysis, where PAS has been proven to be one of the most sensitive optical techniques available [20–22]. In PAS, intensity- or wavelength-modulated light enters the sample cell. If absorption occurs, it leads to modulated temperature changes and consequently to modulated pressure changes (i.e., to the generation of an acoustic wave). Conventional capacitor microphones can be used to detect the sound, but the highest sensitivities have been obtained with special techniques, such as quartz-enhanced PAS (QEPAS) [23–25] and cantilever-enhanced PAS (CEPAS) [23,26–28]. In QEPAS, the frequency of the light-field modulation is tuned to coincide with the sharp resonance of a piezoelectric quartz tuning fork that acts as the sound detector. Because the resonance is narrow, simultaneous detection of multiple down-converted optical frequencies is limited to a narrow optical bandwidth [29]. In contrast, CEPAS is based on a silicon cantilever whose movements induced by the pressure changes inside the sample cell are monitored interferometrically. Highly sensitive detection can be performed outside the resonance frequency of the cantilever, allowing simultaneous detection of multiple down-converted optical frequencies – CEPAS is thus compatible with FTIR [30–36].

A drawback of highly sensitive broadband photoacoustic detection (such as CEPAS) is that the detection bandwidth is limited to low acoustic frequencies (well below 1 kHz), requiring slow scanning velocities with the FTIR instrument. For example, mechanical scanning velocities in the order of 1 mm/s are typically needed in the mid-infrared, which leads to acquisition times ranging from a few seconds to several minutes (depending on the chosen resolution) [33–35]. Here, we demonstrate a 13-fold speed improvement in CEPAS by combining it with the PC-FTS method instead of traditional FTIR. The light source we use is a mid-infrared optical frequency comb (MIR OFC) generated using femtosecond pulse-trapped optical parametric generation with continuous wave seeding [37]. As a proof of concept of the PC-FTS-CEPAS technique, we measured the P-branch of the $v_3$ antisymmetric stretching rovibrational absorption band of methane in the mid-infrared (3.3-3.5 μm) region with a 0.25 Hz scan rate and 10 GHz (0.33 cm$^{-1}$) typical resolution.

In the following, we first explain the experimental setup, after which we focus on the benefits and drawbacks of the PC-FTS method itself in more detail. We then present the results of the broadband methane measurements and compare the performance of the background-free CEPAS detection to a conventional transmission spectroscopy approach. The Supplementary Notes contain additional information about the group- and phase-delay corrections required in PC-FTS, the MIR OFC light source, and other important details of our experimental setup.

## 2. Experimental setup

The PC-FTS-CEPAS setup is depicted in Fig. 1. The PC-FTS part is basically a Michelson interferometer similar to that used in traditional FTIR. However, instead of a translating mirror, the delay arm has a rotating mirror, a focusing optic, and a grating in 4f geometry [5]. The 4f geometry ensures back-reflection despite the deflection caused by the mirror tilt. The grating disperses the input light onto the rotating mirror surface, creating a larger phase-delay difference between adjacent frequencies than in traditional FTIR where all the frequencies would strike the same spot on the mirror. This results in a large group delay and therefore a high resolution with only a small tilt of the mirror. The rotating mirror we used is a simple 1-inch (25.4 mm) square silver mirror (Thorlabs ME1S-P01) mounted on a galvanometric scanner (GSI VM500+). The optics in the reference arm mimic the geometry of the delay arm. This ensures similar beam propagation in the two arms and efficient overlap and interference of the combined beams on the detector side. The most relevant components for PC-FTS (such as the grating groove density and the focal length in the 4f geometry) are discussed in Section 3 and listed in Table 1.

The half-wave plate in Fig. 1 was used to adjust the polarization of the input light to be perpendicular to the grating grooves (i.e., parallel to the plane of the optical table, the plane of paper). This minimized losses at the grating and resulted in the strongest interference signal (Supplementary Note 4). For the light sources, we used a continuous wave (CW) difference frequency generation (DFG) source [38] (for the group- and phase-delay corrections discussed in Section 4) and the MIR OFC [37] (for broadband spectroscopy). The MIR OFC is described in more detail in Supplementary Note 6 and its spectrum is shown in Fig. 2. Typically, 1–2 mW of the CW DFG power and 24 mW (approx. 0.11 mW/cm$^{-1}$ spectral power density) of the MIR OFC power entered the PC-FTS system. However, only a small fraction of these input powers actually arrives at the CEPAS cell (2 % and 7 % from the delay and reference arms, respectively). High losses are inherent to PC-FTS (see Supplementary Note 4), which is problematic for photoacoustic spectroscopy, where the signal strength is proportional to the optical power entering the sample cell.

The irises before and after the PC-FTS system were used to ensure that the beams from all the light sources travel the same path in the system. Flip mirrors were used to switch between the light sources. The flip mirror shown in Fig. 1 was used to guide the CW light to a wavelength meter (EXFO WA-1500) to perform the group- and phase-delay corrections. The combined beams exiting the interferometer were focused through the 10 cm long CEPAS cell (Gasera Ltd. PA201). The CEPAS cell was equipped with uncoated $CaF_2$ windows and contained either 1 % methane in nitrogen for the absorption measurements, or lab air for background measurements. The measurements were performed at 1000 mbar pressure (measured with a pressure sensor at the outlet of the cell; Honeywell HSCMRNN015PAAA5) and 295 K temperature (lab temperature; temperature of the cell was not actively stabilized). Flushing and gas exchange were performed using a vacuum pump (KNF NMP830KNDC). The cell inlet and outlet were closed during the measurements.

In addition to CEPAS detection, interferograms were simultaneously measured with a MIR detector (Vigo UM-I-10.6) after the CEPAS cell. This way we could obtain the MIR OFC power spectrum for normalizing the CEPAS spectra. We also wanted to compare the absorption spectra measured with these two detection schemes. The interferograms were measured by driving the galvanometric scanner with a ramp signal from a waveform generator (Agilent 33220A). The Transistor-Transistor Logic (TTL) signal of the waveform generator was used to trigger a DAQ (GaGe CompuScope 14200) for interferogram digitization.

The interferograms were digitally high-pass (50 Hz) and low-pass (1 kHz) filtered and further processed in Matlab. This data processing included averaging of interferograms or spectra, the phase- and group-delay corrections, rectangular apodization the interferograms to ensure symmetric trimming about the centerbursts, zero-padding, Fourier-transforming the interferograms into the corresponding spectra, and redefining the frequency axes according to the PC-FTS theory. Note that all the interferograms were double-sided and that the spectra were calculated as the modulus of the complex Fourier-transform. Further processing of the raw spectra depends on the detection method (CEPAS vs. MIR detector) and is explained in Section 5.

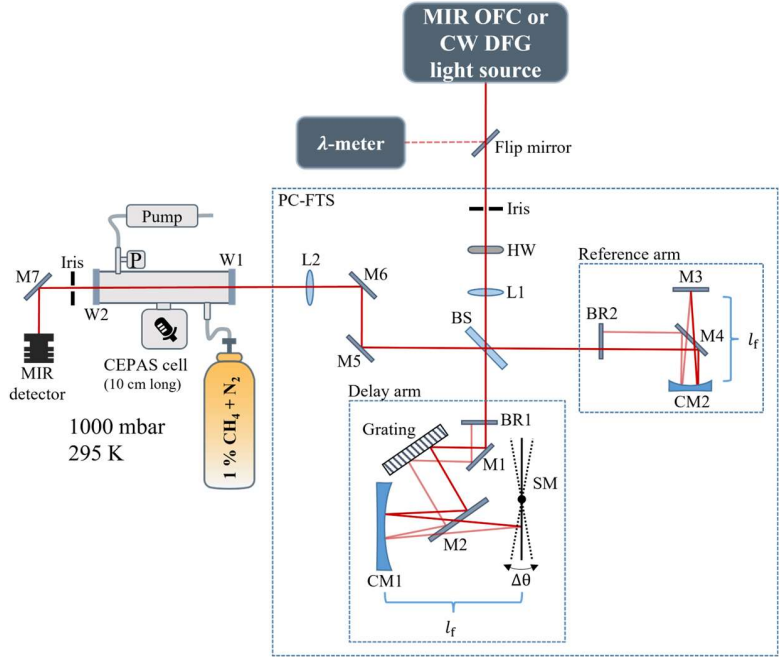

Fig. 1. Experimental setup. M: plane mirror. The mirror materials are given in Supplementary Note 4 to estimate the overall losses; HW: half-wave plate; L: uncoated $CaF_2$ lens; BR: back-reflecting plane mirrors. The lengths of the reference and delay arms are both approx. 83 cm (distance from the BS through the 4f geometry to the BR). Note that the light beams are vertically deflected in both arms such that the BRs reside below the incident beams; CM: 150 mm focal length ($l_f$) curved mirror. The distance from the grating to CM1 is also $l_f$; SM: scanning mirror; $\Delta\theta$: maximum mechanical scan angle; W: CEPAS cell windows (uncoated $CaF_2$); P: pressure sensor; Note that, for simplicity, only a single frequency light beam is drawn to traverse the experimental setup in the schematic. The grating will disperse a broadband light source onto the scanning mirror surface as illustrated in Fig. 1 a of Ref. [5].

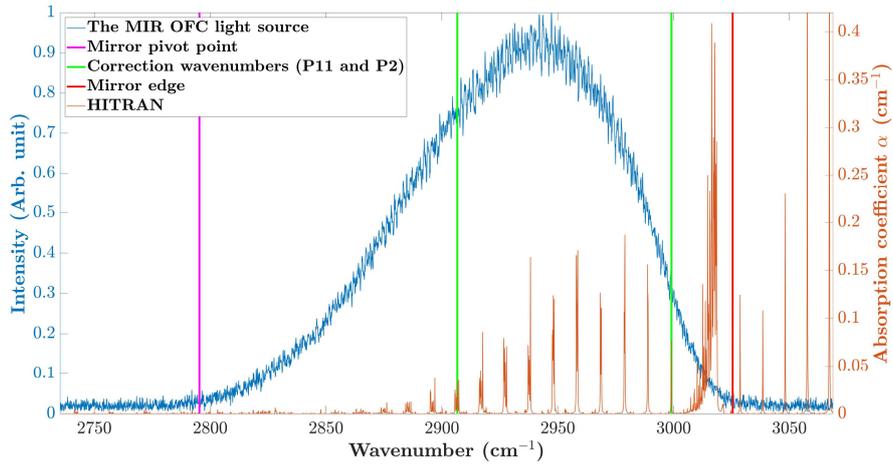

Fig. 2. The MIR OFC spectrum measured with a spectrum analyzer (Bristol 771B; blue trace). The magenta line indicates the optical frequency striking the pivot point of the scanning mirror; the red line indicates the edge of the mirror. The green lines indicate the locations of the methane absorption lines P11 and P2 that were typically used for the phase- and group-delay corrections (Section 4). The orange trace is a simulated methane absorption spectrum (without instrumental broadening) according to the HITRAN database [39,40].

## 3. PC-FTS

In PC-FTS, the optical frequencies are down-converted into corresponding radio frequencies according to Eq. (1). See Supplementary Information of the original paper by Hashimoto and Ideguchi [5] for the derivation.

$$f_{\text{RF}} = c_g (\nu - \nu_0),  \quad (1)$$

where the down-conversion factor

$$c_g = \frac{4 l_f N}{\nu_0} \omega = \frac{8 l_f N \Delta \theta}{\nu_0} f_{\text{scan}}. \quad (2)$$

Here, $l_f$ is the focal length of the focusing optic in the 4f geometry, $N$ is the groove density of the grating, $\nu_0$ is the optical frequency that strikes the pivot point of the scanning mirror and $\omega$ is the angular speed of the scanning mirror. We have expressed the angular speed as $\omega = \Delta\theta/\Delta T = 2\Delta\theta f_{\text{scan}}$, where $\Delta\theta$ is the maximum *mechanical* scan angle, $\Delta T$ is the length of the interferogram in seconds, and $f_{\text{scan}}$ is the scan frequency of the rotating mirror. Note that with this definition, two interferograms are obtained during $1/f_{\text{scan}}$ but in different scan directions (hence the factor of two). For simplicity, we usually omitted the other scan direction.

Eq. (1) reveals that by adjusting the location of the pivot point of the scanning mirror one can choose which optical frequency is mapped into the zero radio frequency. As pointed out by Hashimoto and Ideguchi [5], this freedom of choice makes PC-FTS resemble dual-comb spectroscopy [41] and enables fundamentally faster acquisitions compared to traditional FTIR, where it is the zero optical frequency that is inevitably mapped to the zero radio frequency. As PC-FTS down-converts the optical bandwidth of interest to efficiently fill the whole available RF bandwidth, a lower resolution in the down-converted spectrum (i.e., a shorter measurement time) suffices to resolve the spectral features as illustrated in Fig. 3. In fact (Supplementary Note 1), when the scan frequency and amplitude of the rotating mirror in PC-FTS are matched to the translational speed of the moving mirror in traditional FTIR such that the maximum RFs in the down-converted spectra are the same, the time needed to measure an interferogram with given resolution is reduced by a factor of $\nu_{\max}/\Delta\nu_g$, where $\Delta\nu_g = \nu_{\max} - \nu_0$ is the optical bandwidth down-converted by the PC-FTS method. For example, our rotating mirror supports an optical bandwidth $\Delta\nu_g = 6.9$ THz (230 cm$^{-1}$) with $\nu_{\max} = 90.7$ THz (3025.5 cm$^{-1}$). The speed improvement in our case is thus 13-fold compared to traditional FTIR. However, note the trade-off between the speed benefit and the optical bandwidth; Supplementary Note 2 discusses how to predict the optical bandwidth with chosen scanning mirror width and 4f geometry components. Fig. 2 illustrates the excellent accommodation of the MIR OFC spectrum onto the wide scanning mirror we used.

We define optical resolution as the full width at half maximum (FWHM) of the instrument lineshape function (ILS) when measuring a double-sided interferogram and using rectangular apodization (in which case the ILS is a sinc function). The optical resolution is then

$$\delta\nu = \frac{1}{\Delta T c_g} \times 2 \times 0.603, \quad (3)$$

where $\Delta T$ is the length of the interferogram in seconds and $c_g$ is the down conversion factor. The factor of two considers the double-sidedness and the factor 0.603 casts the base of the sinc function into the FWHM [1]. Note that Hashimoto et al. [5,6] typically measured one-sided interferograms and defined resolution as the base of the sinc function (i.e., without the factor 0.603).

The lengths of the longest interferograms we measured were $\Delta T = 1.82$ s and a typical down-conversion factor was $c_g = 9.1505 \times 10^{-11}$, which down-converted the optical bandwidth with the chosen 4f geometry components (Table 1) below approximately 630 Hz.

This corresponds to $\delta\nu$ = 7.2 GHz (0.24 cm$^{-1}$), which is the highest (double-sided) optical resolution obtainable with our setup. For comparison, in traditional FTIR such resolution would require a 2.5-cm mechanical scan length. Because the FTIR down-conversion factor is $2u/c$, where $u$ is the mechanical scan velocity and $c$ the speed of light [1], $u$ = 1.04 mm/s would be required to down-convert the maximum optical frequency of interest ($\nu_{max}$ = 90.7 THz or 3025.5 cm$^{-1}$) to the 630 Hz value. This corresponds to a recording time of 2.5 cm/$u$ = 24 s, which agrees with the speed-benefit discussion above.

The resolution of our PC-FTS setup is limited by the width of the grating. As the angle of the rotating mirror is scanned, resolution improves until the beam misses the optical components in the 4f geometry (Supplementary Note 3). We also chose to use a slightly poorer resolution (lower scan amplitude) than the highest obtainable value for the broadband measurements and used rectangular apodization to ensure symmetrical trimming about the centerbursts of the interferograms. The expected resolution was then reduced to 8 GHz. However, the experimentally observed resolution was typically somewhat poorer than this due to self-apodizing effects. We characterized the resolution by measuring the spectrum of our monochromatic but wavelength-tunable CW DFG light source at different optical frequencies. The FWHMs of these spectra ranged from 8.7–9.4 GHz (9.0 GHz mean) with the MIR detector and 9.5–11.7 GHz (10.4 GHz mean) with CEPAS detection. In particular, the CEPAS interferograms suffered from self-apodizing effects due to irregular mirror scanning: the response of the CEPAS detector is not constant with respect to the down-converted radio frequency, for which reason nonlinear scan velocity couples to the signal strength. The experimental characterization of the instrument lineshape function is discussed in length in Supplementary Note 8.

Table 1 lists the specifications of the 4f geometry components we used and the PC-FTS performance we obtained. For comparison, we have also included in Table 1 the specifications of the PC-FTS setup Hashimoto et al. used for MIR gas phase spectroscopy with two different spectral resolutions [6]. The main difference between their setup and ours is the different interferogram acquisition rates: they demonstrated high-speed measurements at 12 kHz scan rate whereas we require lower scan rates for our application. In their case, the resolution was not limited by the widths of the gratings they used (52 mm) [6]. For this reason, the grating width and the scan angles are left empty in Table 1. The change from low resolution (2.9×0.603 cm$^{-1}$) to high (0.29×0.603 cm$^{-1}$) required them to limit the optical bandwidth from 300 cm$^{-1}$ to approximately 63 cm$^{-1}$. We can accommodate high optical bandwidth simultaneously with high resolution due to the wide rotating mirror we used.

We have further collected lookup tables in Supplementary Note 10. They list typical components (such as $N$ and $l_f$) available for different spectral ranges (such as MIR and NIR) and the performance expected with them (such as speed benefit and resolution). We hope these tables are useful for anyone wishing to implement PC-FTS for their application.

**Table 1. Specifications of the PC-FTS setup used in this work and in Ref. [6] (two different resolutions).**

| Parameter[a] | This work | Ref. [6] low resolution | Ref. [6] high resolution |
|---|---|---|---|
| $\tilde{\nu}_0$/cm$^{-1}$ | 2795.273[b] | 2464[c,d] | 2204[c,d] |
| $\Delta\tilde{\nu}$/cm$^{-1}$ | 230 | 317[d] | 63[d] |
| $N$/mm$^{-1}$ | 300 | 40 | 180 |
| $l_\text{f}$/mm | 150 | 150 | 150 |
| $\delta\nu$ | 7.2 GHz (0.24 cm$^{-1}$)[e] | 54 GHz (3.0×0.603 cm$^{-1}$) | 5.2 GHz (0.29×0.603 cm$^{-1}$)[f] |
| $W$/mm | 12.7 | 3.6[g] | 3.6[g] |
| $D$/mm | 25 | | |
| $\Delta\theta$ | 4.8°[e] | | |
| $f_\text{scan}$ | 0.25 Hz | 12 kHz | 12 kHz |
| $F$ | 13 | 8 | 35 |

[a]Symbol explanations. $\tilde{\nu}_0$: wavenumber striking the pivot point of the scanning mirror; $\Delta\tilde{\nu}$: optical bandwidth supported by the scanning mirror width; $N$: groove density of the grating; $l_\text{f}$: focal length of the focusing optic in the 4f geometry; $\delta\nu$: optical resolution; $W$: width of the scanning mirror (pivot point to the mirror edge). Limits the optical bandwidth; $D$: width of the grating by which the resolution is limited in our setup (Supplementary Note 3); $\Delta\theta$: estimated maximum scan angle; $f_\text{scan}$: frequency of the mirror scanning; $F$ the speed benefit factor (maximum optical frequency divided by the optical bandwidth).

[b]A slightly different value compared to the broadband methane measurements discussed in Section 5 due to different phase- and group-delay corrections.

[c]Note that in Ref. [6] the PC-FTS system was designed such that it is the highest optical frequency that strikes the mirror pivot point.

[d]Personal communication with Dr. K. Hashimoto

[e]A slightly lower resolution (smaller $\Delta\theta$) was used for the broadband methane measurements discussed in Section 5.

[f]Obtained with one-sided measurements. For comparison, their maximum group delay (120 ps) would correspond to 10.1 GHz (0.335 cm$^{-1}$) resolution if double-sided interferograms were measured and if our definition for resolution was used (FWHM of a sinc function).

[g]7.2×5.0 mm ellipse-shaped mirror [6]

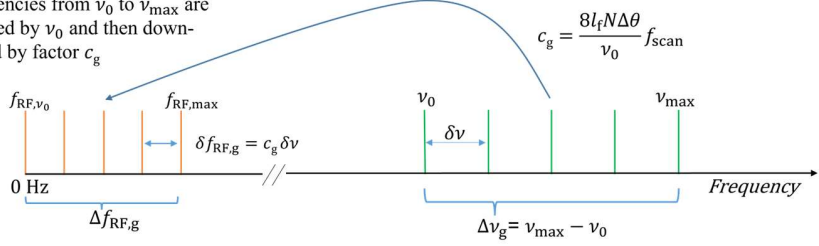

PC-FTS:
All frequencies from $\nu_0$ to $\nu_{max}$ are first shifted by $\nu_0$ and then down-converted by factor $c_g$

$$c_g = \frac{8l_f N \Delta\theta}{\nu_0} f_{scan}$$

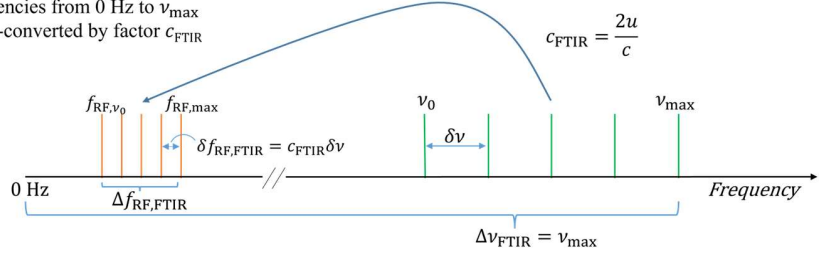

FTIR:
All frequencies from 0 Hz to $\nu_{max}$ are down-converted by factor $c_{FTIR}$

$$c_{FTIR} = \frac{2u}{c}$$

Fig. 3. Illustration of the difference between the down-conversion principles of PC-FTS and traditional FTIR. The PC-FTS approach down-converts the optical bandwidth of interest ($\Delta\nu_g$) to efficiently fill the whole detection bandwidth (0 Hz – $f_{RF,max}$), which leads to fundamentally faster acquisitions, as lower resolution in the down-converted spectrum (i.e., shorter interferogram) suffices to resolve the optical components. See Table 1 for the explanation of the symbols in the PC-FTS down-conversion factor $c_g$; the symbol $u$ in the FTIR down-conversion factor $c_{FTIR}$ denotes the mechanical scan velocity of a traditional FTIR instrument. The symbol $c$ denotes the speed of light. Note that a frequency comb light source is not a prerequisite for PC-FTS, but for simplicity the optical spectrum is drawn as discrete lines to illustrate the optical resolution.

## 4. Phase and group delay correction

A traditional FTIR instrument is typically equipped with a HeNe laser that serves as an inner frequency reference. Its interferogram is monitored simultaneously with the interferogram of the broadband light source and a data point is acquired every time the HeNe interferogram crosses zero. This ensures the broadband interferogram is sampled linearly with respect to the group delay. A similar kind of group-delay (GD) correction is required also in PC-FTS. In addition to GD correction, a special type of phase-delay (PD) correction is required because the scanning mirror is not infinitely thin (the mirror surface usually does not lie on the pivot point). It follows that even the frequency $\nu_0$ experiences phase delay when the mirror is scanned. This must be corrected to recover the spectra accurately [5].

Note that this *phase-delay correction* is not the same as the *phase correction* [1,2] that is typically used in FTIR to reduce noise, process one-sided interferograms and to accurately recover the spectra as the real part of the complex Fourier transform. Here, we do not phase correct the spectra but measure two-sided interferograms and present the spectra as the modulus of the complex Fourier-transform. This is due to the emissive nature of the background-free CEPAS measurements: the phase spectrum is not well defined where signal intensity is zero [1].

The GD and PD corrections are described in detail in Supplementary Note 7. Briefly, the CW DFG light source is used to measure CW interferograms at two different optical frequencies. These CW interferograms are used to calculate the group-delay curve that is then used to resample the interferogram at constant intervals of the group delay. The CW interferograms are also used to calculate the phase-delay curve for a frequency that is the least delayed during scan (i.e., for $\nu_0$). This phase delay-curve is then used to induce a phase shift prior to resampling the interferogram with respect to the linearized group delay and performing the Fourier transform. Note that we define the down-conversion factor as the slope of the linearized group delay. The down-conversion factor and the $\nu_0$ value can be used to switch between the optical and down-converted frequency axes according to Eq. (1).

In principle, any two frequencies can be used for the corrections, but we found that a large wavenumber separation is preferable. In addition, to measure CW interferograms with CEPAS, an absorbing sample is required inside the sample cell and the CW light source wavenumber must be tuned onto an absorption line. We typically used the large wavenumber separation of 92 cm$^{-1}$ between the methane absorption lines P11 and P2 as indicated in Fig. 2. In addition, we used separate GD and PD corrections for the interferograms measured with the MIR detector and with CEPAS detection (correction curves determined for the MIR detector could not be readily used for CEPAS and vice versa). It is noteworthy that in Ref. [6] the corrections were performed using a single CW laser, which is possible if the rotating mirror is placed on a translational stage and the missing wavenumber reference is taken from an absorption feature in a measured broadband spectrum (see the Supplementary Information of Ref. [6]).

As the PD and GD corrections required us to measure CW interferograms at two distinct optical frequencies, they were not performed in real time. This means that the GD and PD correction curves are determined based on a separate measurement, and the curves are stored, and used in future measurements to correct the interferograms. It may be preferable to perform the GD and PD correction measurements on the same day as the actual broadband measurements, but there was no conclusive inaccuracy if correction curves of a separate day were used. However, the lack of real-time correction did pose problems with residual phase noise as discussed in Supplementary Note 8. The phase noise distorts the broadband spectra as discussed in the following section.

## 5. Broadband absorption spectroscopy

Fig. 4 shows typical group-delay corrected broadband interferograms from a background (lab air) measurement and a methane measurement using the MIR detector (left panel), and from a methane measurement using the CEPAS detector (right panel). Each interferogram is an average of 10 interferograms. Note the fundamental differences between the interferograms measured with the two detection methods. The centerburst of an interferogram measured with the MIR detector mainly describes the envelope of the light source, and any structure in the spectrum (such as methane absorption) is encoded into the sidebursts further away from the centerburst. In contrast, PAS is a background-free technique, which means that ideally all structure in the interferogram is due to absorption. This leads to efficient use of the dynamic range of the detector and foreshadows higher signal-to-noise ratios (SNRs) in the PAS spectra than in the MIR detector spectra [1].

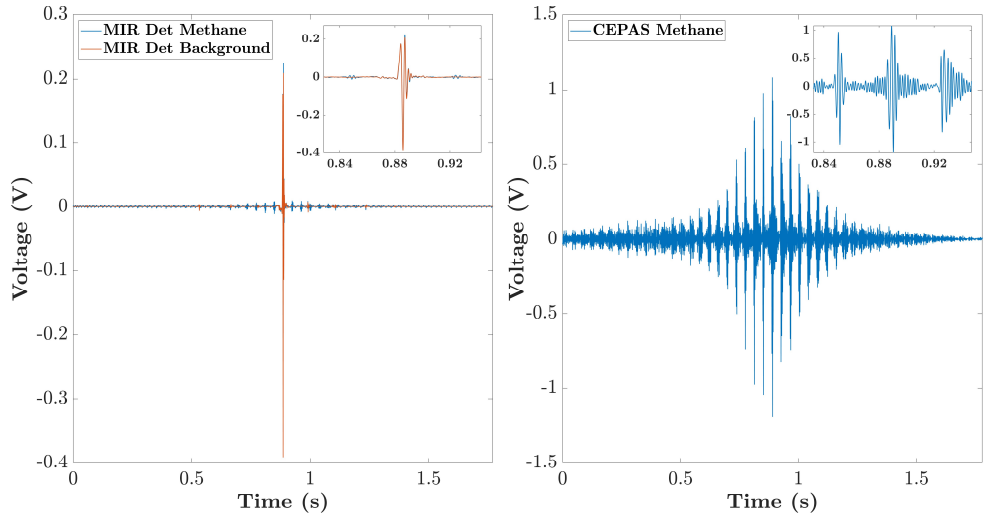

Fig. 4. Typical interferograms measured with the MIR detector (left panel) and CEPAS (right panel). The lengths of the time windows are 1.77 s and 1.78 s for the MIR detector interferograms and the CEPAS interferogram, respectively. The corresponding down-conversion factors are $8.5027\times10^{-11}$ and $8.4887\times10^{-11}$, respectively. These values differ due to the different phase- and group-delay corrections used for the two detection methods.

Four interferograms like the ones shown in Fig. 4 (MIR detector background, MIR detector methane, or CEPAS methane; each an average of 10 interferograms) were processed into a spectrum and averaged. Without averaging, the envelopes of subsequently measured spectra varied from spectrum to spectrum and prevented the determination of a reliable transmission spectrum. The variation from spectrum to spectrum is presumably due to the jitter discussed further in Supplementary Note 8; the mirror scans were not perfectly reproducible from scan to scan, which leads to inaccurate phase- and group-delay corrections.

Fig. 5 shows the averaged spectra based on the MIR detector measurements. To yield the absorption spectrum based on these raw spectra, the methane spectrum was divided by the background spectrum, after which the Beer–Lambert law [42] was applied to obtain the absorption coefficient $\alpha = -\ln(I/I_0)/L$, where $I$ is the methane spectrum, $I_0$ the background spectrum, and $L$ the absorption path length. Division by the background spectrum left a residual sloping, which was corrected using Savitzky–Golay filtering. In addition, a constant offset of 1.9 GHz was applied to the frequency axis. The resulting final absorption spectrum based on the MIR detector measurements is shown in Fig. 6. Only the methane lines P14–P3 are shown, as lines beyond this range were masked by noise.

Fig. 7 shows the raw spectrum based on the CEPAS measurements. To yield the absorption spectrum, the raw spectrum was normalized with the MIR detector background spectrum, as the PAS signal at each optical frequency is proportional to the optical power at that frequency. In addition, the CEPAS spectrum was normalized with the CEPAS response plotted as the yellow line in Fig. 7 (the response was determined experimentally, see Supplementary Note 9). Furthermore, CEPAS requires calibration with respect to the sample concentration. This was considered by fitting a scaling parameter to yield a good match between the measured spectrum and the chosen HITRAN reference discussed below. Finally, a residual background was subtracted using Savitzky–Golay filtering, and a constant shift of 1.8 GHz was applied to the frequency axis. The resulting final absorption spectrum based on the CEPAS measurements is shown in Fig. 8.

The black dotted line in the final MIR detector spectrum in Fig. 6 (CEPAS spectrum in Fig. 8) is a simulation of the methane spectrum via simulating the interferogram of a HITRAN transmission (absorption) spectrum, multiplying it with the experimental apodization function and by reprocessing the interferogram into an absorption spectrum. The experimental apodization function was determined from the average envelope of a few CW interferograms measured at lines P11, P6, and P2 (see discussion in Supplementary Note 8). It can be seen that the measured methane lines are much flatter and wider than expected. The orange line in Fig. 6 (Fig. 8) is a convolution of a HITRAN transmission (absorption) spectrum with a Lorentzian function with its FWHM fixed to 9 GHz (10.5 GHz) [39,40]. This provided the best match with the measured spectrum compared to a few different convolution functions we tried. The HITRAN references assume air-broadened lines even though the methane measurements were performed in nitrogen. In addition, speed-dependent effects on the lineshapes and line-mixing are known to be important for methane [43]. However, inclusion of such effects [43] and changing the air broadenings to nitrogen broadenings [43,44] had negligible effect on the residuals plotted in the lower panels of Figs. 6 and 8. The found discrepancies between the measured and the expected spectra are presumably due to the phase noise (jitter) that the experimental ILS function ignores. This hypothesis was supported by tests where we introduced jitter to simulated methane interferograms. Other reasons for the discrepancies may include different alignment of the MIR OFC light source through the system compared with the CW light source, unaccounted for aberration effects [6,45], or detector nonlinearity, which is also known to distort spectra in a similar manner [46]. More rigorous analysis of the ILS could be performed based on measurements of isolated absorption lines [47] instead of relying on separate CW measurements as we have done.

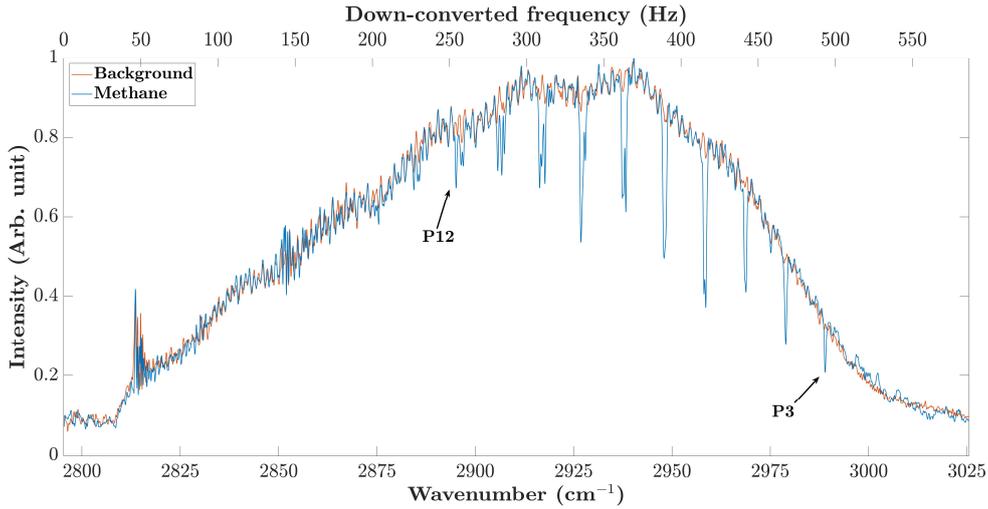

Fig. 5. Raw methane and background spectra measured with the MIR detector. The horizontal axis spans the full bandwidth from the pivot point ($\tilde{\nu}_0 = 2795.372$ cm$^{-1}$) to the edge of the scanning mirror (3025.5 cm$^{-1}$). The artefacts at the down-converted frequencies approx. 45 Hz and 145 Hz are due to the MIR detector fans. The etalon structure is presumably due to the uncoated CaF$_2$ lenses, the CEPAS cell windows, or both. Also note how intensity leaks below 50 Hz despite the high-pass filtering. This leaking is due to the imperfect phase and group delay corrections (residual phase noise) caused by the unreproducible mirror scans (discussed further in Supplementary Note 8).

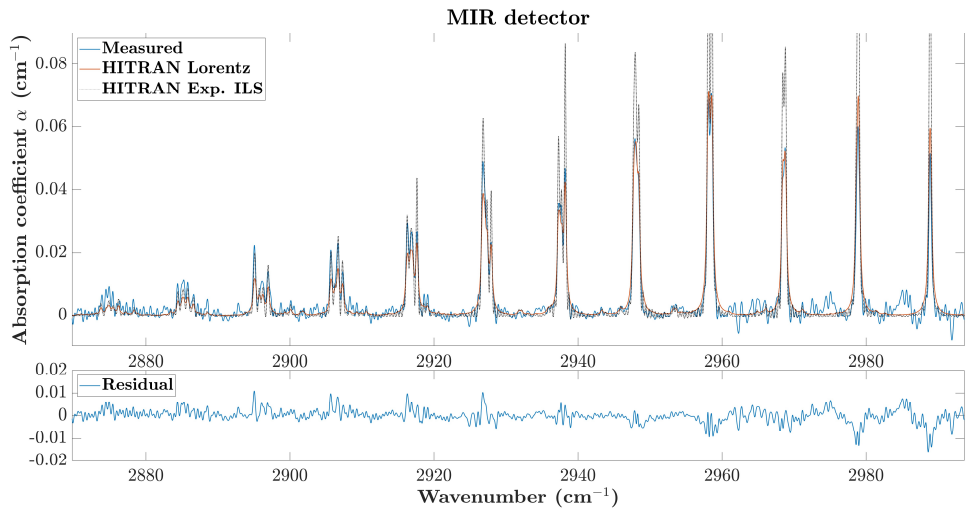

Fig 6. Final absorption spectrum of methane lines P14–P3 measured with the MIR detector (blue trace). Note that especially lines P4 and P3 have poor SNR and uncertainty in the background subtraction. The black dotted trace is a simulation of the expected absorption spectrum. It is the result of simulating the interferogram of a HITRAN transmission spectrum, multiplying it with the experimental apodization function, and reprocessing the result into an absorption spectrum. The orange trace is a HITRAN reference that yields the best match with the measured spectrum. It is the result of convoluting a HITRAN transmission spectrum with a Lorentz function whose FWHM was fixed to 9 GHz. The residual in the lower panel is calculated against the Lorentz reference.

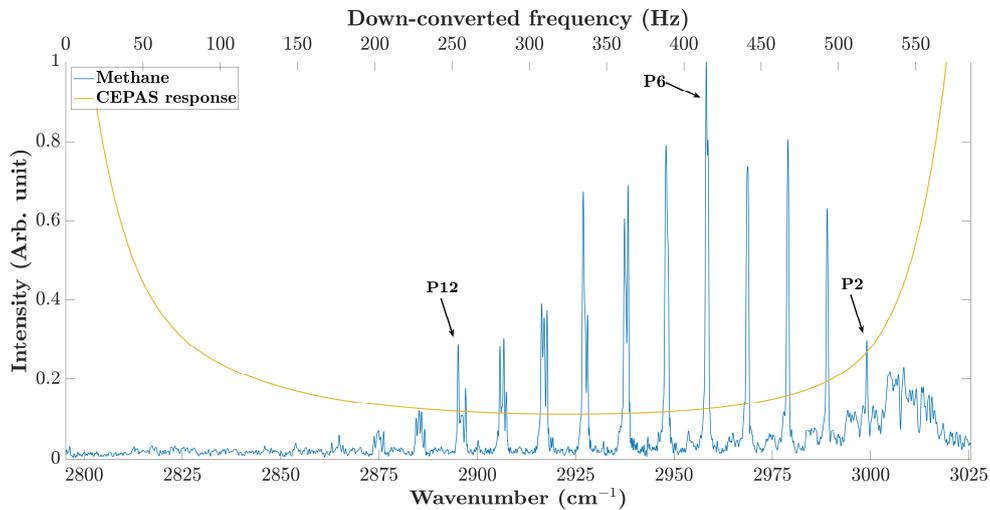

Fig. 7. Raw methane spectrum measured with the CEPAS detector (before normalizing with the MIR detector background spectrum in Fig. 5, or with the CEPAS response plotted as the yellow trace in this figure). The horizontal axis spans the full bandwidth from the pivot point ($\tilde{\nu}_0 = 2795.340$ cm$^{-1}$) to the edge of the scanning mirror (3025.5 cm$^{-1}$). Note the increasing background signal above approx. 3000 cm$^{-1}$ due to noise close to the CEPAS detector resonance (570 Hz). Also note that the P2 line is clearly visible with CEPAS detection but it is masked by noise in the MIR detector measurements (Fig. 5).

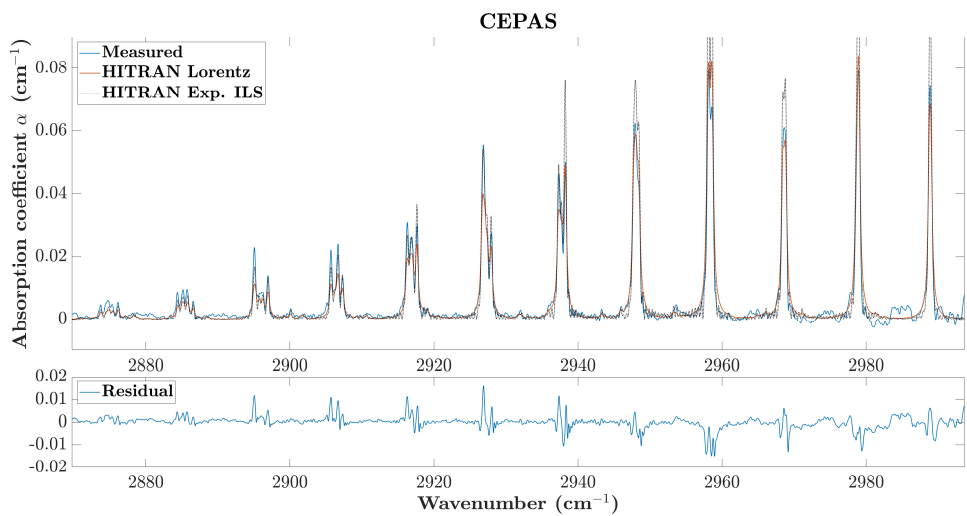

Fig. 8. Final absorption spectrum of methane lines P14–P3 measured with the CEPAS detector (blue trace). The black dotted trace is a simulation of the expected absorption spectrum. It is the result of simulating the interferogram of a HITRAN absorption spectrum, multiplying it with the experimental apodization function and Fourier-transforming the result back into an absorption spectrum. The orange trace is a HITRAN reference that yields the best match with the measured spectrum. It is the result of convoluting a HITRAN transmission spectrum with a Lorentz function whose FWHM was fixed to 10.5 GHz. The residual in the lower panel is calculated against the Lorentz reference.

The standard deviations of the MIR detector spectrum residuals (Fig. 6) and the CEPAS spectrum residuals (Fig. 8) are both approximately 4 % from the respective maximum values in the spectra (the line P6 for both). As the residuals under the peaks mainly describe the match of the measured spectrum with the assumed reference spectrum, we define SNR as the line P6 height divided by the standard deviation of the residual with the residuals under the absorption peaks and data above approximately 2970 cm$^{-1}$ excluded. This approach yields an SNR of 57 for the MIR detector spectrum and of 116 for the CEPAS spectrum. For the MIR detector spectrum, the SNR corresponds to a noise-equivalent limit of detection (NELOD) of 180 ppm.

The NELOD for the CEPAS spectrum should be determined in a different way than for the MIR detector spectrum. Mikkonen et al. [35] have shown that PAS detection leads to noise reduction in FTIR compared with traditional detection schemes, as only the spectral components absorbed by the sample contribute to the observed noise. Similarly, it is clear that the phase-noise contribution by a spectral component (Supplementary Note 8) scales with the absorbed power. This means that as the sample concentration is decreased, the absorption peaks and the noise decrease with constant SNR until the background noise of the detector starts to dominate. The NELOD should then be determined from a separate CEPAS spectrum measurement where light enters the sample cell but no methane is present. We measured such a spectrum (a single spectrum from an average of 10 interferograms), calculated the standard deviation of the noise between 2870 cm$^{-1}$ and 2970 cm$^{-1}$, divided the P6 peak height of the raw methane spectrum in Fig. 7 with the calculated standard deviation, and arrived at a NELOD of 7 ppm. This is a factor of 25 better than the NELOD obtained with the MIR detector. The NELOD for CEPAS is of the same order or in some cases a factor of ten poorer than values reported in previous CEPAS-FTS demonstrations [28,33].

## 6. Conclusions

We have demonstrated a 13-fold speed improvement in broadband cantilever-enhanced photoacoustic spectroscopy (CEPAS) by using the phase-controlled Fourier-transform spectroscopy (PC-FTS) instead of traditional Fourier-transform infrared spectroscopy (FTIR). Indeed, the typical 1.8 s long interferograms measured with PC-FTS would require up to 24 seconds of scanning to yield the expected optical resolution of 7.2 GHz with traditional FTIR. In practice, the resolution was degraded due to self-apodization effects caused by non-constant scan velocity of the rotating mirror. In addition, unreproducible scanning resulted in residual phase noise, which distorted the broadband spectra. These problems could potentially be avoided by using a more stable scanner, implementing the phase- and group-delay corrections in real time [48,49], or both. The optical resolution could then be improved by using a wider grating, unless aberration effects prevent this [6,45]. It is noteworthy that the speed benefit of PC-FTS is a trade-off with the optical bandwidth. The scanning mirror we used accommodated a maximum bandwidth of 6.9 THz. However, we were able to consider only the P-branch of methane due to the limited tuning of the MIR OFC light source. Despite the inherently high optical losses of PC-FTS and the power scaling of CEPAS signal, the CEPAS detection was shown to be at least 25 times more sensitive than the traditional transmission spectroscopy approach.

**Funding.** Academy of Finland (326444); S. Larnimaa acknowledges financial support from the CHEMS doctoral program of the University of Helsinki.

**Acknowledgments.** We thank Dr. Kazuki Hashimoto, Prof. Takuro Ideguchi and Dr. Markus Metsälä for their insights and constructive feedback on the project.

Author contributions: S. Larnimaa constructed the experimental setup (not including the mid-infrared optical frequency comb light source), performed the measurements, analyzed the data, and wrote the manuscript; M. Roiz constructed the mid-infrared optical frequency comb light source, assisted with its use, and reviewed the manuscript; M. Vainio conceptualized and supervised the project, and reviewed the manuscript.

**Disclosures.** The authors declare no conflicts of interest.

**Data availability.** Data and Matlab codes underlying the results presented in this paper are not publicly available at this time but may be obtained from the authors upon reasonable request.

**Supplemental document.** See Supplement 1 for supporting content.

# Photoacoustic phase-controlled Fourier-transform infrared spectroscopy: supplemental document

**This document is organized as follows:**

- Supplementary Note 1 (SNote1): The speed benefit factor of PC-FTS is derived.
- SNote2: The optical bandwidth limit with given 4f geometry components is estimated.
- SNote3: The maximum resolution limited by the grating width is explained.
- SNote4: The losses in the PC-FTS system are discussed.
- SNote5: The expected interference contrast with the given losses is shown.
- SNote6: The mid-infrared optical frequency comb light source is briefly reviewed.
- SNote7: The phase- and group-delay corrections are illustrated.
- SNote8: The continuous wave interferogram measurements and the experimental instrument lineshape function are discussed in detail.
- SNote9: The coupling of nonconstant velocity scanning to the CEPAS signal strength is illustrated by an interferogram simulation.
- SNote10: lookup tables estimating the PC-FTS performance at different optical frequency ranges and with given 4f geometry components are gathered.

**SNote1. Speed benefit of PC-FTS**

Let the time it takes to measure an interferogram be $\Delta T$. The resolution in the down-converted radio frequency (RF) spectrum is then $1/\Delta T$ (see Eq. (3) of the main text; here we omit the extra factor 2×0.603 for simplicity). The optical resolution $\delta\nu$ is deduced from the RF resolution by dividing it by the down conversion factor. We denote the down-conversion factor in PC-FTS by $c_g$, and in traditional FTIR by $c_{FTIR}$. The subscript g refers to "galvanometric scanner" and is used to keep the notation consistent with the notation in the original paper by Hashimoto and Ideguchi [1].

One way to express the down-conversion factor is to calculate the ratio of the down-converted RF bandwidth $\Delta f_{RF} = (f_{RF,max} - 0)$ to the optical bandwidth $\Delta \nu$. In PC-FTS, the optical bandwidth to be down-converted is given by $(\nu_{max} - \nu_0)$, where $\nu_{max}$ is the optical frequency that hits the edge of the scanning mirror and is mapped to $f_{RF,max}$, and $\nu_0$ is the optical frequency that hits the pivot point and is mapped to the zero radio frequency. In traditional FTIR, it is the zero optical frequency that is inevitably mapped to the zero radio frequency, i.e, the optical bandwidth is simply $\nu_{max}$. All in all, the times it takes to measure an interferogram with PC-FTS ($\Delta T_g$) and with traditional FTIR ($\Delta T_{FTIR}$) with given optical resolution are:

$$\Delta T_g = \frac{1}{\delta\nu c_g} = \frac{\Delta\nu_g}{\delta\nu \Delta f_{RF}} = \frac{\nu_{max} - \nu_0}{\delta\nu f_{RF,max}} \quad (S1)$$

$$\Delta T_{FTIR} = \frac{1}{\delta\nu c_{FTIR}} = \frac{\Delta\nu_{FTIR}}{\delta\nu \Delta f_{RF}} = \frac{\nu_{max}}{\delta\nu f_{RF,max}} \ . \quad (S2)$$

When the scan frequency and amplitude of the rotating mirror in PC-FTS are matched to the translational speed of the moving mirror in traditional FTIR such that the maximum RFs in the down-converted spectra are the same, we find that PC-FTS interferograms can be recorded, with fixed optical resolution, by a factor of

$$\frac{\Delta T_{\text{FTIR}}}{\Delta T_{\text{g}}} = \frac{\nu_{\max}}{\Delta \nu_{\text{g}}} \tag{S3}$$

faster than in traditional FTIR.

**SNote2. Optical bandwidth**

The speed benefit of PC-FTS is in trade-off with the optical bandwidth, which is limited by the width of the scanning mirror. The width of the diverged spot $W$ on the mirror surface can be deduced using the grating equation [2]:

$$\varphi_{\text{m}} = \arcsin(Nm\lambda - \sin(\varphi_{\text{i}})) \, , \tag{S4}$$

where $\varphi_{\text{m}}$ is the diffraction angle, $N$ is the groove density, $m$ is the diffraction order, $\lambda$ is the wavelength and $\varphi_{\text{i}}$ is the incidence angle. In the following, we will assume the diffraction order $m = 1$. The angles are measured against the grating surface normal, and they are positive if they are on the same side of the normal.

According to Fig. S1, the required mirror width $W$ to accommodate a bandwidth of $\lambda_0 - \lambda_{\max}$ is

$$W = 2l_{\text{f}} \times \tan\left(\frac{\Delta\varphi}{2}\right) , \tag{S5}$$

where $\Delta\varphi = \varphi_{\lambda_0} - \varphi_{\lambda_{\max}}$. The full equation is then

$$W = 2l_{\text{f}} \times \tan\left(\frac{\arcsin(N\lambda_0 - \sin(\varphi_{\text{i}})) - \arcsin(N\lambda_{\max} - \sin(\varphi_{\text{i}}))}{2}\right) \tag{S6}$$

$$\approx 2l_{\text{f}} \times \tan\left(\frac{N\lambda_0 - \sin(\varphi_{\text{i}}) - [N\lambda_{\max} - \sin(\varphi_{\text{i}})]}{2}\right) = 2l_{\text{f}} \times \tan\left(\frac{N[\lambda_0 - \lambda_{\max}]}{2}\right)$$

$$\approx l_{\text{f}} N \Delta\lambda = l_{\text{f}} N c \frac{\Delta\nu}{\nu_0 \nu_{\max}} \, . \tag{S7}$$

The inaccuracy of the approximate equation (S7) depends on the grating groove density, the wavelength, the optical bandwidth and the incidence angle on the grating (we used 65° incident angle in our setup). Although the equation nicely illustrates the dependencies between the different design parameters, use of the full equation (S6) for designing your PC-FTS setup is recommended.

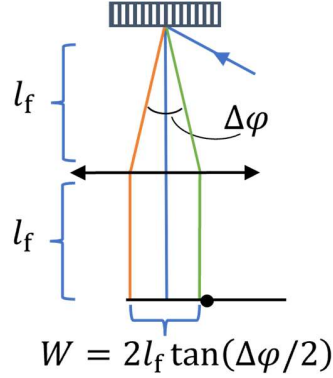

Fig. S1. Illustration of the required approximate scanning mirror width $W$ in terms of the diffraction angle span $\Delta\varphi$ that depends on the grating groove density, incidence angle and the optical bandwidth.

The scanning mirror we used is a 1 inch (25.4 mm) square mirror. It is important to note that, in a typical case where the pivot point is located in the middle of the mirror, *only half of the mirror width can be used*, i.e., the optical bandwidth needs to be accommodated between the pivot point and the edge of the mirror. In our case, the usable mirror width is then $W = 12.7$ mm. Any optical frequency that falls on the other side of the pivot point leads to aliasing. To ensure accommodation, one may need to use optical filtering, or use a grating with a lower groove density with the expense of lower resolution.

In our case, no optical filtering was required as the optical bandwidth of the MIR optical frequency comb was an excellent fit with the wide scanning mirror we chose, as illustrated in Fig. 2 of the main text. We found experimentally that, with the chosen pivot point location ($\tilde{\nu}_0 = 2795$ cm$^{-1}$), the highest wavenumber supported by the mirror was $\tilde{\nu}_{\max} = 3025.5$ cm$^{-1}$. With these inputs (and the rest given in Table 1 of the main text), Eq. (S7) yields a 12.3 mm spot on the mirror.

## SNote3. Maximum resolution

When the theoretical down-conversion factor given by Eq. (2) of the main text is inserted into Eq. (3) of the main text, one can predict the resolution that can be obtained with given 4f geometry components:

$$\delta\nu = \frac{\nu_0}{4Nl_f\Delta\theta} \times 2 \times 0.603 \,. \tag{S8}$$

The resolution enhancement of PC-FTS is due to the grating. The higher the groove density $N$, the more two adjacent optical frequencies are displaced on the scanning mirror and the smaller scan angle is required to resolve them. According to Eq. (S8), using a larger focal length focusing optic ($l_f$) would appear to have the same effect: the longer the focal length, the more the two frequencies diverge relative to one another before collimation by the focusing optic. However, the *maximum* resolution for a given groove density is limited by the largest allowed beam walk-off during a scan. In practice, this means that for a given focal length and scan angle the grating needs to be at least $D = 2l_f\Delta\theta$ wide as illustrated in Fig. S2. In addition, the focusing optic needs to be at least $2l_f\Delta\theta + W$ wide ($W$ is the distance between the pivot point and the edge of the rotating mirror). The curved mirror we used as the focusing optic has a diameter of 2 in. (50.8 mm). In addition, the width of the mirror denoted by M2 in Fig. 1 of the main text is 36 mm. The back-reflecting mirror (BR1) and the mirror before the grating (M1) were both 1 in. wide.

In our setup the resolution is limited by the width of the grating. In that case, the maximum resolution is given by

$$\delta\nu_{\max} \approx \frac{\nu_0}{2ND} \times 2 \times 0.603 \,. \tag{S9}$$

Eq. (S9) shows that the focal length and scan angle are in trade-off in terms of the resolution: the larger the focal length, the less one can scan the angle of the rotating mirror before the beam misses the optical components and leaves the overall resolution unchanged. We used $l_f = 150$ mm. This choice was further motivated by the fact that a 100 mm focal length would not have been feasible due to tight spacing of the optics in the delay arm, and that a 200 mm focal length would not have increased resolution based on the discussion above but reduced the optical bandwidth. In addition, note that increasing $l_f$ increases the lengths of the delay and reference arms four times the change in $l_f$.

We used a grating of width $D = 2.5$ cm. With $D = 2l_f\Delta\theta$, we can estimate the maximum mechanical scan angle: $\Delta\theta \approx 4.8° = \pm 2.4°$. With these values (and the rest given in Table 1 of the main text) we get from Eq. (S9) that the theoretical maximum resolution for our setup is 6.7 GHz. This is close to the maximum resolution we actually obtained (7.2 GHz).

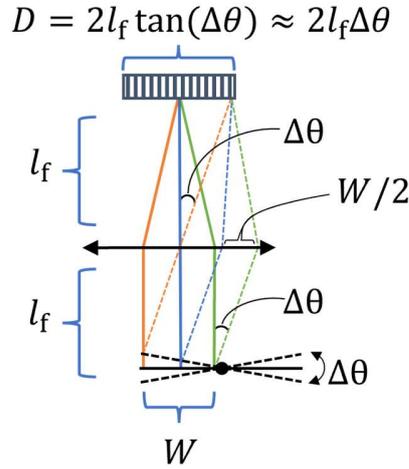

Fig. S2. Illustration of the required approximate widths of the grating ($D$) and the focusing optic ($D + W$) for a given maximum scan angle $\Delta\theta$ when the optical bandwidth is accommodated onto the scanning mirror of width $W$.

## SNote4. Losses

A drawback of PC-FTS is high losses due to the 4f geometry: the beam propagating in the delay arm strikes the grating a total of four times. The grating we used (Dynasil ML-601) has the specified efficiency of 86 % for the S-polarization (perpendicular to the grating grooves, i.e., in the plane of the optical table) at 3.4 µm wavelength. In practice the measured efficiency was 76 %. The total efficiency due to the grating is then 33.4 %.

We used a beam splitter with 50/50 split ratio for unpolarized light (Thorlabs BSW511). However, the strongest interference signal was obtained by having P-polarized light (perpendicular to the reflective surfaces, i.e., in the plane of the optical table; note the different definition for the grating) adjusted with the half-wave plate (see Fig. 1 of the main text). This minimized losses at the grating but resulted in a largely unbalanced split ratio and maximized the power in the delay arm (79/17 transmission $T$ to reflectivity $R$ ratio). This is a nonideality as the best beam splitter efficiency (product $R \times T$) is obtained with 50/50 split ratio.

The rest of the components (the HW-plate, the lenses, and the collection of different mirrors) are expected to have a total efficiency of 56.8 % for the beam taking the delay arm path and 60.8 % for the reference arm (estimated at 3.4 µm wavelength based on the efficiency graphs provided by the component manufacturer: Thorlabs). Overall, this means that of the original intensity entering the PC-FTS system (measured between the flip mirror and the first iris in Fig. 1 of the main text) only 2.5 % is expected before the CEPAS cell from the delay arm and 8.2 % from the reference arm (measured between L2 and W1). The measured efficiency before the CEPAS cell was 1.8 % and 7 % for the delay and reference arms, respectively. Further reductions in the power reaching the detectors are expected due to the uncoated $CaF_2$ windows (94.4 % efficiency for each) of the CEPAS cell and the mirror before the MIR detector (97.9 %). The full component list with the estimated efficiencies is given in Tables S1 and S2.

As mentioned in the main text, typically 1–2 mW of the DFG light source and 24 mW of the MIR OFC entered the PC-FTS system. Despite the fact that only a small fraction of these powers arrives at the CEPAS cell, we obtained strong CEPAS signals with the 1 % sample concentration. The signal levels of the interferograms measured with the MIR detector were only modest. In addition, despite the efficiency mismatch of the arms, the interference contrast is expected to be 92 % for a CW interferogram and 85 % for a broadband interferogram (see SNote5).

**Table S1. Delay arm component efficiencies estimated at the 3.4 μm wavelength based on the efficiency graphs provided by the manufacturer (Thorlabs).**

| Component[a] | Description/material | Efficiency (%) |
|---|---|---|
| HW | Half-wave plate | 99.4 |
| L1 | Uncoated $CaF_2$ | 94.4 |
| BS | Transmission | 79[b] |
| M1 | Gold Enh. 45° | 97.9 |
| Grating | Grating | 76[b] |
| M2 | Silver 45° | 97.9 |
| CM1 | Gold 0° | 96.5 |
| SM | Silver 0° | 96.4 |
| CM1 | Gold 0° | 96.5 |
| M2 | Silver 45° | 97.9 |
| Grating | Grating | 76[b] |
| M1 | Gold Enh. 45° | 97.9 |
| BR1 | Aluminum 0° | 98.4 |
| M1 × Grating × M2 × CM1 × SM × CM1 × M2 × Grating × M1 | Repeat | 97.9×76[b]×97.9× 96.5×96.4×96.5× 97.9×76[b]×97.9=47.6 |
| BS | Reflection | 17[b] |
| M5 × M6 | 2 × Gold Enh. 45° | 97.9×97.9=95.8 |
| L2 | Uncoated $CaF_2$ | 94.4 |
| | **Total** | **2.5** |

[a]Not included in the table are the uncoated $CaF_2$ windows of the CEPAS cell (W1 and W2 in Fig. 1 of the main text; 94.4 % efficiency per window) and the MIR enhanced gold mirror before the MIR detector (M7; 97.9 %). Note that the tables have multiple entries for the same components. HW: half-wave plate; L: lens; BS: beam splitter; M: plane mirror; CM: curved mirror; the angle values refer to the angle of incidence; Gold Enh.: MIR enhanced gold; Repeat: the beam interacts with previously listed components.

[b]Experimentally measured value.

**Table S1. Reference arm efficiencies estimated at the 3.4 μm wavelength based on the efficiency graphs provided by the manufacturer (Thorlabs).**

| Component[a] | Description/material | Efficiency (%) |
|---|---|---|
| HW | Half-wave plate | 99.4 |
| L1 | Uncoated $CaF_2$ | 94.4 |
| BS | Reflection | 17[b] |
| M4 | Silver 45° | 97.9 |
| CM2 × M3 × CM2 | 3 × Gold 0° | 96.5×96.5×96.5=89.9 |
| M4 | Silver 45° | 97.9 |
| BR2 | Silver 0° | 96.4 |
| M4 × CM2 × M3 × CM2 × M4 | Repeat | 97.9× 89.9× 97.9=86.2 |
| BS | Transmission | 79[b] |
| M5 × M6 | 2 × Gold enh. 45° | 97.9×97.9=95.8 |
| L2 | Uncoated $CaF_2$ | 94.4 |
| | **Total** | **8.2** |

[a]Not included in the table are the uncoated $CaF_2$ windows of the CEPAS cell (W1 and W2 in Fig. 1 of the main text; 94.4 % efficiency per window) and the MIR enhanced gold mirror before the MIR detector (M7; 97.9 %). Note that the table has multiple entries for the same components. HW: half-wave plate; L: lens; BS: beam splitter; M: plane mirror; CM: curved mirror; the angle values refer to the angle of incidence; Gold Enh.: MIR enhanced gold; Repeat: the beam interacts with previously listed components.

[b]Experimentally measured value.

## SNote5. Interference contrast

The interference signal from a CW light source at the detector is given by [3]

$$I(\phi) = I_0 RT(\eta_{\text{ref}} + \eta_{\text{del}}) + 2I_0 RT \sqrt{\eta_{\text{ref}} \eta_{\text{del}}} \cos(\phi) \ , \tag{S10}$$

where $I_0$ is the intensity entering the PC-FTS system, $R$ is the reflectivity and $T$ the transmission of the beam splitter, $\eta_{\text{ref}}$ is the total efficiency of the components for the beam travelling the reference arm path and $\eta_{\text{del}}$ for the delay arm, and $\phi$ is the phase delay induced by scanning the delay arm. With $T = 0.79$, $R = 0.17$, $RT\eta_{\text{ref}} = 0.082$ (Table S2) and $RT\eta_{\text{del}} = 0.025$ (Table S1), we can estimate the expected interference contrast (Eq. (S11)) in a CW interferogram to be 92 %. In practice, before high-pass filtering the detector signals, an interference contrast as high as 80 % was observed with the MIR detector for the P11 CW interferogram presented in Fig. S4.

$$\frac{I(0) - I(\pi)}{I(0)} = \frac{4\sqrt{\eta_{\text{ref}} \eta_{\text{del}}}}{\eta_{\text{ref}} + \eta_{\text{del}} + 2\sqrt{\eta_{\text{ref}} \eta_{\text{del}}}} \tag{S11}$$

For broadband interferograms, the interference contrast can be defined as

$$\frac{I(0) - I(\infty)}{I(\infty)} = \frac{2\sqrt{\eta_{\text{ref}} \eta_{\text{del}}}}{\eta_{\text{ref}} + \eta_{\text{del}}} \ , \tag{S12}$$

where $I(\infty) = I_0 RT(\eta_{\text{ref}} + \eta_{\text{del}})$ is basically the DC intensity on top of which the interferogram lies. The interference strength is then expected to be 85 %. Before high-pass filtering of the detector signal, the broadband MIR detector interferograms presented in Fig. 4 of the main text typically lay on top of a 0.8 V DC level making the interference strength only 25 %. This discrepancy is unknown.

## SNote6. The MIR OFC light source

The broadband light source we used is a novel MIR optical frequency comb (MIR OFC) developed in our lab [4–6]. It is based on single-pass femtosecond pulse-trapped optical parametric generation seeded with a CW laser [7]. Briefly [4], femtosecond pulses from a 1.04 µm optical frequency comb (the pump) together with a beam from a 1.5 µm CW laser (the seed) are guided through a nonlinear crystal (MgO doped periodically poled lithium niobate). From the pump pulses, signal pulses in the 1.5 µm and idler pulses in the 3.4 µm optical regions are generated and efficiently amplified thanks to the pulse trapping effect [7]. The purpose of the seed is to lower the threshold of the nonlinear process, improve the pulse-to-pulse coherence of the MIR OFC and reduce the relative intensity noise. Particularly, the seeding allows a simple means of fully stabilizing the OFC if part of the pump power is sampled for supercontinuum generation: when the CW laser is phase locked to the supercontinuum, also the carrier envelope offset is locked and freely tunable even if the carrier envelope offset frequency of the pump comb is free running. The repetition rate (frequency separation of adjacent comb teeth) of the MIR OFC is defined by the pump comb repetition rate (250 MHz), which is also tunable.

The spectral envelope of the MIR OFC can be tuned between 3325–4000 nm (2380–3030 cm$^{-1}$) with a typical FWHM of 100 nm [4]. The tuning is carried out by first choosing the poling period of the fanout nonlinear crystal. Then, the seed wavelength is tuned until the idler output power is maximized. Finally, the pump power can be tuned to adjust the idler power. The nonlinear process is most efficient at higher idler wavelengths, and the average idler power can be as high as 700 mW. Here, however, we operate at the low wavelength tuning limit of the MIR OFC. We used 1556.4 nm seeding (with 20 mW seed power; high seed powers reduce the threshold and hence maximize the idler power) and 4.4 W pump power. These settings produced 100 mW of average idler power and a spectrum shown in Fig. 2 of the main text. It can be seen that the MIR OFC envelope covers only the P-branch of methane. Trying to push the tunability of the comb resulted in a considerably lower efficiency of the process (i.e., low idler powers) without considerably increasing the spectral coverage towards the R-branch. Coverage of the R-branch could be attained by adjusting the pump central wavelength to adjust the optimal pulse trapping conditions.

The comb spectrum shown in Fig. 2 of the main text has its maximum at 88.2 THz (3399 nm; 2942 cm$^{-1}$) with 88 nm FWHM (113 cm$^{-1}$; 3.39 THz). Note the etalon structure, which presumably originates from an uncoated $CaF_2$ lens used to collimate the OFC beam after the nonlinear crystal. Of the 100 mW average output power available, only 24 mW was typically let into the PC-FTS system. Power was attenuated using a device with two silicon plates in Brewster angles (Altechna). Power attenuation was necessary to limit the CEPAS signals obtained with the 1 % methane concentration within the voltage input limits of our data acquisition card (±2 V). From Fig. 2 of the main text, one can estimate that the base width of the OFC is approximately 6.5 THz or 216.8 cm$^{-1}$ (calculated as range 84–90.5 THz or 2802–3018.8 cm$^{-1}$). With the 250 MHz repetition rate of the comb, the average power per comb tooth is therefore approximately 1 µW, or 0.11 mW/cm$^{-1}$.

## SNote7. Phase- and group-delay correction

The phase and group delay (PD and GD) corrections were performed by measuring CW interferograms at two different wavenumbers. We typically used interferograms measured at the methane absorption lines P11 and P2. In this specific example, the wavenumbers are 2906.684 cm$^{-1}$ and 2999.105 cm$^{-1}$, respectively. The top left panel of Fig. S3 shows the interferograms (after high-pass and low-pass filtering). They represent for all practical purposes the highest resolution obtainable with our current setup: the signal clips at the start and end of the scan as the beam misses the grating. The interferograms are then trimmed and the zero crossings are determined (top right panel of Fig. S3). Every zero crossing advances the phase by $\pi$, and thus we obtain the phase delay curves for the two optical frequencies, plotted in the bottom left panel of Fig. S3. The PD curves are then used to determine the group delay (plotted in the bottom right panel, right axis), which is the frequency derivative of the phase delay [1]:

$$\tau_g(t) = c_g(t)t = -\frac{1}{2\pi}\frac{\partial \phi_\nu(t)}{\partial \nu} \equiv -\frac{1}{2\pi}\frac{\phi_{\nu 2}(t) - \phi_{\nu 1}(t)}{\nu_2 - \nu_1}, \quad (S13)$$

where $\phi_{\nu x}(t) = -2\pi c_g(t)(\nu_x - \nu_0(t))t$ is the phase delay for the optical frequency $\nu_x$ and $c_g(t)$ is the down-conversion factor, which is time dependent in the usual case where the mirror velocity is not perfectly constant during a scan. Here, $\nu_0(t)$ indicates that all frequencies experience a common delay if the scanning mirror surface does not lie on the pivot point.

The green curve in the bottom right panel of Fig. S3 (left axis) is the phase delay curve for a frequency close to the pivot point of the scanning mirror. Ideally, if the scanning mirror was infinitely thin, this optical frequency $\nu_0$ would not experience any phase delay during the scan. Note that the choice of the constant $\nu_0$ is somewhat arbitrary. We define $\nu_0$ as the frequency for which the phase returns to zero at the end of the scan. It is the optical frequency whose down-converted, linearized radio-frequency is zero. The phase delay for the chosen constant $\nu_0$ is then determined using the phase delays of the two other frequencies by calculating

$$\phi_{\nu 0}(t) = (x+1)\phi_{\nu 2}(t) - x\phi_{\nu 1}(t)$$
$$= (x+1)\bigl(-2\pi[\nu_2 - \nu_0(t)]\tau_g(t)\bigr) - x\bigl(-2\pi[\nu_1 - \nu_0(t)]\tau_g(t)\bigr)$$
$$= -2\pi\bigl([(x+1)\nu_2 - x\nu_1] - \nu_0(t)\bigr)\tau_g(t) \equiv -2\pi\bigl(\nu_0 - \nu_0(t)\bigr)\tau_g(t), \quad (S14)$$

where

$$\nu_0 = [(x+1)\nu_2 - x\nu_1] \Leftrightarrow x = \frac{\nu_0 - \nu_2}{\nu_2 - \nu_1}. \quad (S15)$$

To PD and GD correct a measured interferogram, it is first multiplied with [8,9]

$$e^{-i\phi_{\nu 0}(t)}, \quad (S16)$$

after which the interferogram is resampled to be linear with respect to the group delay $\tau_g(t)$, and finally Fourier-transformed. Throughout the text we refer with the down-conversion factor $c_g$ to the slope of the linearized group delay curve. The down-conversion factor and the $\nu_0$ value can be used to switch between the optical and down-converted radio frequency axes according to Eq. (1) of the main text.

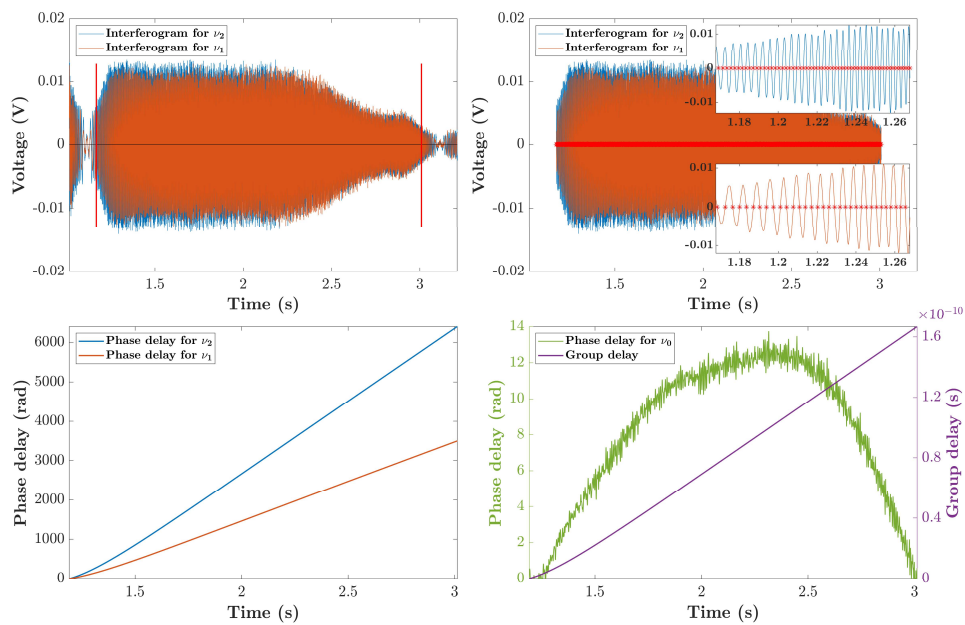

Fig. S3. Illustration of the phase and group delay corrections. Top left panel: the high-pass and low-pass filtered raw CW interferograms at 2906.684 cm$^{-1}$ ($\nu_1$) and 2999.105 cm$^{-1}$ ($\nu_2$). The red lines indicate the chosen trimming points for the interferograms; Top right panel: the trimmed interferograms with their zero crossings determined (the insets show zoomed-in views of the beginnings of the interferograms); Bottom left panel: the phase delay curves determined based on the zero crossings; Bottom right panel: the phase delay for the frequency $\nu_0$ that strikes the pivot point of the scanning mirror (green curve, left axis). The violet curve (right axis) is the group delay determined based on the phase delay curves of the optical frequencies $\nu_1$ and $\nu_2$. The maximum group delay is 166.7 ps.

## SNote8. CW interferograms and spectra

To study the accuracy of the phase-delay (PD) and group-delay (GD) corrections, inspect the realization of the expected resolution (the shape of the instrument lineshape function) and to detect other nonidealities in the PC-FTS performance, we measured continuous wave (CW, i.e., single frequency) interferograms with the DFG light source. Fig. S4 shows typical CW interferograms (each is an average of 10 interferograms after high-pass and low-pass filtering) at two different wavenumbers (the methane absorption lines P11 and P2 at 2906.684 cm$^{-1}$ and 2999.105 cm$^{-1}$ typically used for the corrections). It can be seen that, for the MIR detector measurements, the shapes of the interferograms are practically the same for both wavenumbers. On the contrary, the CEPAS measurements show more variation – the P2 measurement is a drastic example of the behavior of the CEPAS interferograms. This behavior is explained by the fact that the response of the CEPAS detector is not constant as a function of the down-converted radio frequency (RF). Even though the scanning mirror was driven with a ramp signal, its velocity and therefore the RF was not constant during the scan. For example, in the case of the P2 interferogram, the RF sweeps over the resonance (570 Hz) of the CEPAS detector evidenced in Fig. S4 by the signal reaching its maximum at around 0.5 s. In SNote9 we reproduce this effect by simulating the CEPAS interferogram based on a separately measured response curve of the CEPAS detector and the instantaneous RF curve determined from the zero-crossings of the corresponding MIR detector interferogram. To improve the quality of the CEPAS interferograms, as linear as possible scanning should be attained and/or the down-converted frequency bandwidth should be accommodated below the resonance and within the flat region of the CEPAS response (see Fig. S7, or Fig. 7 of the main text for the CEPAS response).

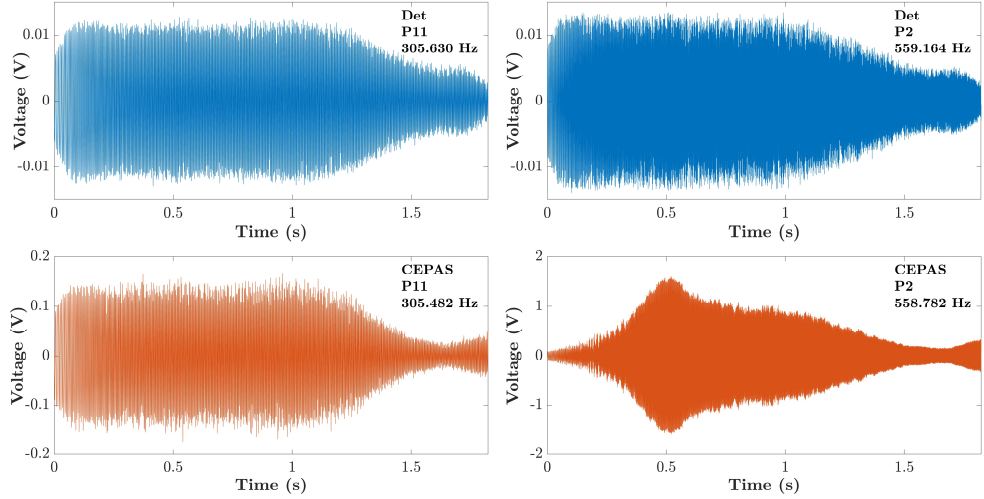

Fig. S4. Typical CW interferograms measured at two different wavenumbers (the methane lines P11 and P2 at 2906.684 cm$^{-1}$ and 2999.105 cm$^{-1}$) with the MIR detector (upper panels) and CEPAS (lower panels). Note that the interferograms at P11 (and similarly the interferograms at P2) correspond to one another, i.e., they have been measured at the same time. The linearized (i.e., average) radio frequencies are shown in the figure insets. The lengths of all the interferograms are approx. 1.82 s. The down-conversion factor for the MIR detector (CEPAS) interferograms is 9.1505×10$^{-11}$ (9.1420×10$^{-11}$) and the $\tilde{v}_0$ is 2795.273 cm$^{-1}$ (2795.223 cm$^{-1}$). The values are slightly different for the two detection methods due to slightly different data trimming and due to that separate PD and GD corrections are used for the two detection methods.

Fig. S5 shows the corresponding spectra of the MIR detector interferograms of Fig. S4. In addition, a CW spectrum at the P6 transition (2958.081 cm$^{-1}$) is shown. The spectra are obtained by PD and GD correcting the interferograms and Fourier-transforming them. Note that the P11 and P2 interferograms of Fig. S4 are exactly the ones used for the corrections; SNote7 shows the PD and GD correction using the MIR detector interferograms. It can be seen from Fig. S5 that the P11 and P2 spectra are beautifully recovered with practically zero offsets from their true wavenumbers, as expected. In contrast, the P6 spectrum shows slight asymmetry, offset from the correct wavenumber and sidelobes of phase noise. This exemplifies that the PD and GD corrections do not work perfectly for interferograms that are not used in the corrections themselves; even for subsequent interferograms measured at P11 or P2, phase noise still emerges. This is explained by the fact that the scan of the mirror is not perfectly reproducible from scan to scan. Indeed, studying the zero crossings of subsequent single interferograms (i.e., before averaging) we found that the standard deviation of corresponding zero-crossings was typically 0.23 ms. As an example, this is as high as 14 % of the period of an RF at 600 Hz and sure to introduce noise into the spectra [10–12]. It is then obvious that to suppress the phase noise and improve the accuracy of the PD and GD corrections, a much more stable scanner should be used, or one should be able to perform the PD and GD corrections in real time.

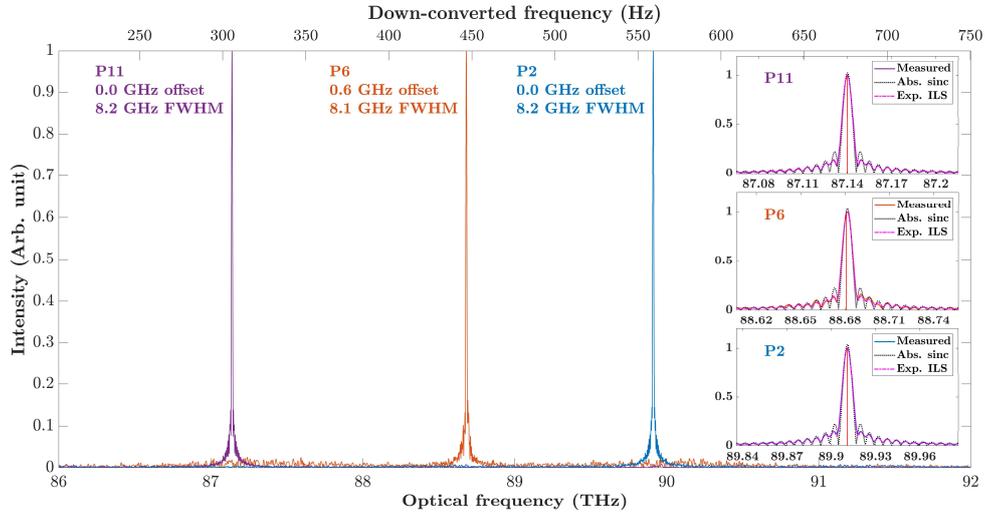

Fig. S5. The corresponding spectra of the MIR detector interferograms in Fig. S4 together with a CW measurement at the methane line P6 (2958.081 cm$^{-1}$). The texts laid on the figure show offsets of the peaks from their true wavenumbers and the FWHMs of the peaks. The insets show zoomed-in views of the peaks. In the insets, the dotted line is a fit of (the absolute value of) a sinc function with the FWHM fixed to the expected resolution (7.2 GHz). The magenta dashed line is the expected instrument lineshape function derived from the average envelopes of all the interferograms. The red lines indicate the true wavenumbers measured with a wavelength meter (to which the P11 and P2 measurements are referenced via the PD and GD corrections). Note the residual phase noise in the P6 spectrum not corrected by the PD and GD corrections. This is in contrast to the P11 and P6 spectra whose corresponding interferograms were exactly the ones used for the corrections.

The expected resolution of the MIR detector CW spectra is 7.2 GHz according to Eq. (3) of the main text, i.e., according to the time window length and the down-conversion factor (given in the caption of Fig. S4). This represents for all practical purposes the highest (double sided) resolution obtainable with our current setup: increasing the scan amplitude did not improve the resolution as the beam already clips at the grating at the start and end of the scan (see SNote7). However, from Fig. S5, one can see that the experimentally observed resolution (FWHMs of the peaks) is approximately 8.2 GHz. This is caused by the self-apodization effect (decline in the interferogram amplitude) evident for the MIR detector interferograms in Fig. S4. This self-apodization is due to an unfortunate alignment issue we could not rectify for the time being. However, the resulting instrument line shape function (ILS) still resembles a sinc function as seen in the insets of Fig. S5, where for each line we have fitted (the absolute value of) a sinc function with the FWHM fixed to the expected resolution. In addition, to simulate the experimental instrument lineshape function, we have plotted the absolute value of a Fourier-transform of a cosine that has been multiplied with the average envelope of all the CW interferograms. It is seen that this model excellently describes all the lines but obviously neglects the phase noise of the P6 measurement. The effect of the phase noise is to distort the broadband spectra as discussed in Section 5 of the main text.

As expected based on the discussion on the shapes of the CEPAS interferograms above, the shape of the ILS of the CEPAS measurements is wavenumber dependent, as seen in Fig. S6. Despite the expected resolution (7.2 GHz), the FWHMs of the lines can vary between 8.8 GHz and 10.3 GHz making it difficult to define a simple reference for the broadband methane spectra discussed in Section 5 of the main text. Note also the increased phase noise close to the CEPAS detector resonance (570 Hz), and the larger offset (2.4 GHz) of the P6 line from the correct wavenumber. A few GHz offsets were typical also for the MIR detector measurements despite the smaller value of the specific example shown in Fig. 5.

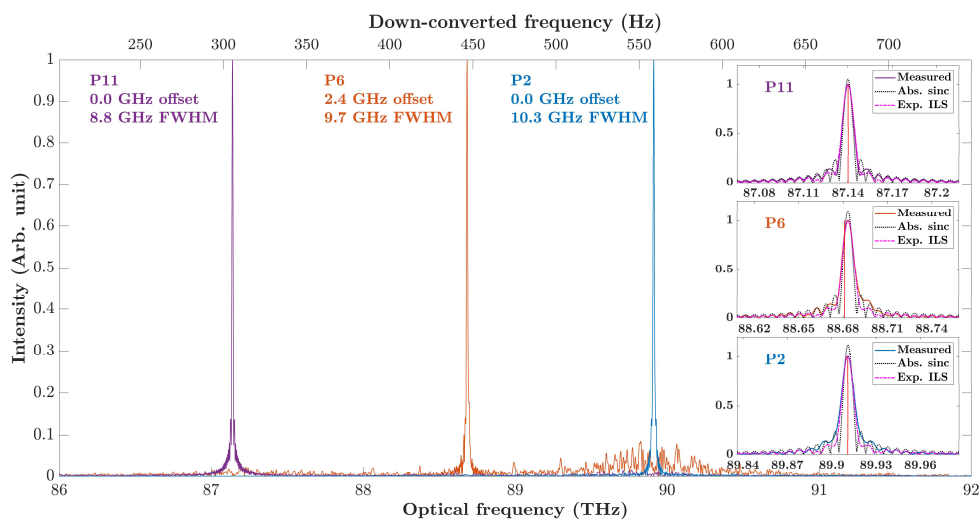

Fig. S6. The corresponding spectra of the CEPAS interferograms of Fig. S4 together with a CW measurement at the methane line P6 (2958.081 cm$^{-1}$). The texts laid on the figure show offsets of the peaks from their true wavenumbers and the FWHMs of the peaks. The insets show zoomed-in views of the peaks. In the insets, the dotted line is a fit of (the absolute value of) a sinc function with the FWHM fixed to the expected resolution (7.2 GHz). The magenta dashed line is the expected ILS derived from the average envelopes of all the interferograms. Note the amplified phase noise of the P6 spectrum close to the CEPAS resonance (570 Hz). Also note the larger variation of the FWHMs compared to the corresponding MIR detector spectra in Fig. S5.

For the broadband measurements discussed in Section 5 of the main text, the scan amplitude was slightly lowered and the broadband interferograms were rectangularly apodized in order to have the centerbursts in the middle of the time windows. These choices lowered the linearized radio frequency of the P2 line (560 Hz → 520 Hz) well below the CEPAS resonance (570 Hz) and lead to 8.0 GHz expected resolution. The same type of CW interferogram analysis as discussed above was performed with the following results: the FWHMs of the CW spectra ranged from 8.7–9.4 GHz (9.0 GHz mean) with the MIR detector measurements and 9.5–11.7 GHz (10.4 GHz mean) with CEPAS.

### SNote9. CEPAS interferogram simulation

The CW interferogram measured with the MIR detector at 2999.105 cm$^{-1}$ is presented in Fig. S7. In addition, the envelope of the interferogram and its instantaneous down-converted radio frequency (calculated based on the inverses of adjacent zero crossings) have been plotted. It is evident that the velocity of the scanning mirror is not constant during the scan but accelerates and sweeps over the resonance of the CEPAS response (plotted in the bottom left panel of Fig. S7). The CEPAS response was determined based on measurements where CW light entered the methane filled CEPAS cell and an optical chopper was used as the intensity modulator: time signals at different modulation frequencies were measured and Fourier-transformed, after which the response at the used modulation frequency was defined as the height of the peak in the corresponding spectrum.

The bottom right panel of Fig. S7 shows a cosine multiplied with the experimental MIR detector interferogram envelope and by the CEPAS response as a function of the (smoothed) instantaneous radio frequency. This simulated CEPAS interferogram very closely resembles the experimental CEPAS interferogram that has been measured simultaneously with the corresponding MIR detector interferogram and plotted in the top right panel. It is thus proved that the weird shapes of the CEPAS interferograms are explained by the scan velocity not being constant during a scan.

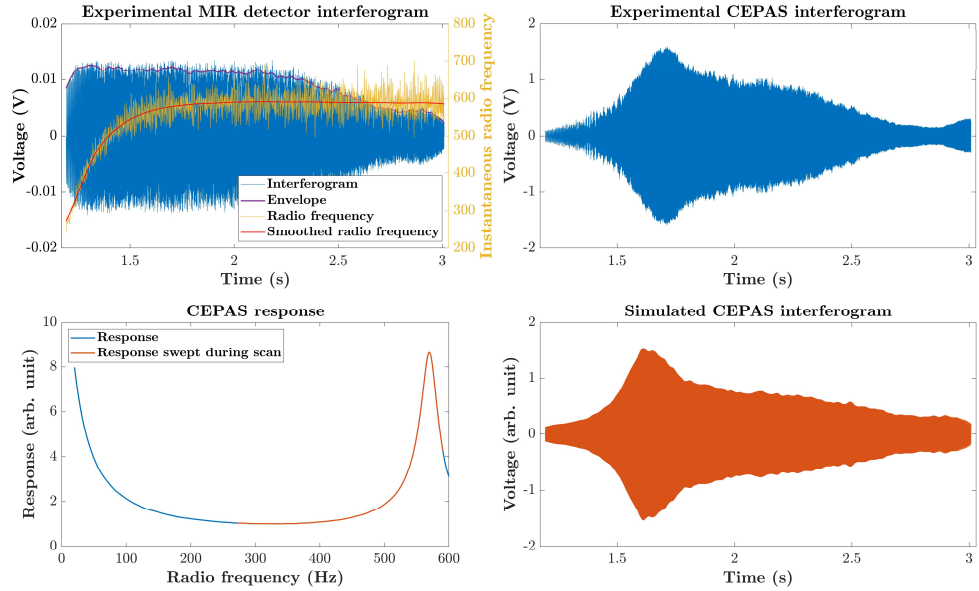

Fig. S7. Top left panel: The experimental MIR detector interferogram at 2999.105 cm$^{-1}$. The violet trace is the envelope of the interferogram. The yellow trace is the instantaneous down-converted radio frequency calculated based on the inverses of adjacent zero crossings. The red trace is the yellow one but smoothed; Top right panel: the experimental CEPAS interferogram simultaneously measured with the MIR detector one; Bottom left panel: the CEPAS response as a function of radio frequency. The orange trace emphasizes the part of the response that is swept during the measurement of the experimental interferograms; Bottom right panel: simulated CEPAS interferogram. It is the result of multiplying a cosine with the experimental MIR detector interferogram envelope and the CEPAS response as a function of the instantaneous radio frequency.

## SNote10. Lookup tables

Tables S3–S7 list expected PC-FTS performance at four different spectral regions (10 μm, 3 μm, 1.5 μm, 1 μm and 0.6 μm) and with different 4f geometry component combinations. The values are only estimates – we do not guarantee their experimental realization [13]. The symbols in the tables are explained as follows:

- $\nu_0$: the optical frequency striking the pivot point of the scanning mirror
- $\Delta\nu$: optical bandwidth
- $N$: groove density of the grating
- $l_f$: focal length of the focusing optic
- $D$: width of the grating that limits the optical resolution according to Eq. (S9)
- $F$: the speed benefit factor (maximum optical frequency divided by the optical bandwidth)
- $\delta\nu$: maximum optical resolution given by Eq. (S9)
- $W$: width of the scanning mirror (distance from the pivot point to the edge) required to accommodate the optical bandwidth according to Eq. (S7)
- $\Delta\theta$: the maximum mechanical scan angle after which the beam misses the grating.

The tables are read by choosing a desired $\Delta\nu$ (the leftmost column), and an $N$, $l_f$ and $D$ combination (the three topmost rows), after which one can read the expected performance with respect to the rest of the parameters.

**Table S2. Lookup table for expected PC-FTS performance at $\lambda_0 = 10$ μm ($\nu_0 = 30$ THz; $\tilde{\nu}_0 = 1000$ cm$^{-1}$). The symbols are explained in the text.**

| $\Delta\nu$ (THz) $\Delta\tilde{\nu}$ (cm$^{-1}$) | $N$ (mm$^{-1}$) | 75 | | | | | | | | | 100 | | | | | | | | |
|---|---|---|---|---|---|---|---|---|---|---|---|---|---|---|---|---|---|---|---|
| | $l_f$ (mm) | 100 | | | 150 | | | 200 | | | 100 | | | 150 | | | 200 | | |
| | $D$ (mm) | 12.5 | 25 | 50 | 12.5 | 25 | 50 | 12.5 | 25 | 50 | 12.5 | 25 | 50 | 12.5 | 25 | 50 | 12.5 | 25 | 50 |
| 2.5 83 | $F$ | 13 | | | | | | | | | 13 | | | | | | | | |
| | $\delta\nu$ (GHz) | 19.3 | 9.7 | 4.8 | 19.3 | 9.7 | 4.8 | 19.3 | 9.7 | 4.8 | 14.5 | 7.2 | 3.6 | 14.5 | 7.2 | 3.6 | 14.5 | 7.2 | 3.6 |
| | $\delta\tilde{\nu}$ (cm$^{-1}$) | 0.64 | 0.32 | 0.16 | 0.64 | 0.32 | 0.16 | 0.64 | 0.32 | 0.16 | 0.48 | 0.24 | 0.12 | 0.48 | 0.24 | 0.12 | 0.48 | 0.24 | 0.12 |
| | $W$ (mm) | 6 | | | 9 | | | 12 | | | 8 | | | 12 | | | 15 | | |
| | $\Delta\theta$ (deg.) | 3.6 | 7.2 | 14.3 | 2.4 | 4.8 | 9.5 | 1.8 | 3.6 | 7.2 | 3.6 | 7.2 | 14.3 | 2.4 | 4.8 | 9.5 | 1.8 | 3.6 | 7.2 |
| 5 150 | $F$ | 7 | | | | | | | | | 7 | | | | | | | | |
| | $\delta\nu$ (GHz) | 19.3 | 9.7 | 4.8 | 19.3 | 9.7 | 4.8 | 19.3 | 9.7 | 4.8 | 14.5 | 7.2 | 3.6 | 14.5 | 7.2 | 3.6 | 14.5 | 7.2 | 3.6 |
| | $\delta\tilde{\nu}$ (cm$^{-1}$) | 0.64 | 0.32 | 0.16 | 0.64 | 0.32 | 0.16 | 0.64 | 0.32 | 0.16 | 0.48 | 0.24 | 0.12 | 0.48 | 0.24 | 0.12 | 0.48 | 0.24 | 0.12 |
| | $W$ (mm) | 11 | | | 16 | | | 21 | | | 14 | | | 21 | | | 29 | | |
| | $\Delta\theta$ (deg.) | 3.6 | 7.2 | 14.3 | 2.4 | 4.8 | 9.5 | 1.8 | 3.6 | 7.2 | 3.6 | 7.2 | 14.3 | 2.4 | 4.8 | 9.5 | 1.8 | 3.6 | 7.2 |
| 10 300 | $F$ | 4 | | | | | | | | | 4 | | | | | | | | |
| | $\delta\nu$ (GHz) | 19.3 | 9.7 | 4.8 | 19.3 | 9.7 | 4.8 | 19.3 | 9.7 | 4.8 | 14.5 | 7.2 | 3.6 | 14.5 | 7.2 | 3.6 | 14.5 | 7.2 | 3.6 |
| | $\delta\tilde{\nu}$ (cm$^{-1}$) | 0.64 | 0.32 | 0.16 | 0.64 | 0.32 | 0.16 | 0.64 | 0.32 | 0.16 | 0.48 | 0.24 | 0.12 | 0.48 | 0.24 | 0.12 | 0.48 | 0.24 | 0.12 |
| | $W$ (mm) | 19 | | | 28 | | | 37 | | | 25 | | | 37 | | | 50 | | |
| | $\Delta\theta$ (deg.) | 3.6 | 7.2 | 14.3 | 2.4 | 4.8 | 9.5 | 1.8 | 3.6 | 7.2 | 3.6 | 7.2 | 14.3 | 2.4 | 4.8 | 9.5 | 1.8 | 3.6 | 7.2 |

Table S4. Lookup table for expected PC-FTS performance at $\lambda_0 = 3$ μm ($\nu_0 = 100$ THz; $\tilde{\nu}_0 = 3333$ cm$^{-1}$). The symbols are explained in the text.

| $\Delta\nu$ (THz) $\Delta\tilde{\nu}$ (cm$^{-1}$) | $N$ (mm$^{-1}$) | 150 | | | | | | | | | 300 | | | | | | | | | 450 | | | | | | | | |
|---|---|---|---|---|---|---|---|---|---|---|---|---|---|---|---|---|---|---|---|---|---|---|---|---|---|---|---|---|
| | $l_f$ (mm) | 100 | | | 150 | | | 200 | | | 100 | | | 150 | | | 200 | | | 100 | | | 150 | | | 200 | | |
| | $D$ (mm) | 12.5 | 25 | 50 | 12.5 | 25 | 50 | 12.5 | 25 | 50 | 12.5 | 25 | 50 | 12.5 | 25 | 50 | 12.5 | 25 | 50 | 12.5 | 25 | 50 | 12.5 | 25 | 50 | 12.5 | 25 | 50 |
| 2.5 83 | $F$ | 41 | | | | | | | | | 41 | | | | | | | | | 41 | | | | | | | | |
| | $\delta\nu$ (GHz) | 32.2 | 16.1 | 8 | 32.2 | 16.1 | 8 | 32.2 | 16.1 | 8 | 16.1 | 8 | 4 | 16.1 | 8 | 4 | 16.1 | 8 | 4 | 10.7 | 5.4 | 2.7 | 10.7 | 5.4 | 2.7 | 10.7 | 5.4 | 2.7 |
| | $\delta\tilde{\nu}$ (cm$^{-1}$) | 1.07 | 0.54 | 0.27 | 1.07 | 0.54 | 0.27 | 1.07 | 0.54 | 0.27 | 0.54 | 0.27 | 0.13 | 0.54 | 0.27 | 0.13 | 0.54 | 0.27 | 0.13 | 0.36 | 0.18 | 0.09 | 0.36 | 0.18 | 0.09 | 0.36 | 0.18 | 0.09 |
| | $W$ (mm) | 1 | | | 2 | | | 2 | | | 2 | | | 3 | | | 4 | | | 3 | | | 5 | | | 7 | | |
| | $\Delta\theta$ (deg.) | 3.6 | 7.2 | 14.3 | 2.4 | 4.8 | 9.5 | 1.8 | 3.6 | 7.2 | 3.6 | 7.2 | 14.3 | 2.4 | 4.8 | 9.5 | 1.8 | 3.6 | 7.2 | 3.6 | 7.2 | 14.3 | 2.4 | 4.8 | 9.5 | 1.8 | 3.6 | 7.7 |
| 5 150 | $F$ | 21 | | | | | | | | | 21 | | | | | | | | | 21 | | | | | | | | |
| | $\delta\nu$ (GHz) | 32.2 | 16.1 | 8 | 32.2 | 16.1 | 8 | 32.2 | 16.1 | 8 | 16.1 | 8 | 4 | 16.1 | 8 | 4 | 16.1 | 8 | 4 | 10.7 | 5.4 | 2.7 | 10.7 | 5.4 | 2.7 | 10.7 | 5.4 | 2.7 |
| | $\delta\tilde{\nu}$ (cm$^{-1}$) | 1.07 | 0.54 | 0.27 | 1.07 | 0.54 | 0.27 | 1.07 | 0.54 | 0.27 | 0.54 | 0.27 | 0.13 | 0.54 | 0.27 | 0.13 | 0.54 | 0.27 | 0.13 | 0.36 | 0.18 | 0.09 | 0.36 | 0.18 | 0.09 | 0.36 | 0.18 | 0.09 |
| | $W$ (mm) | 2 | | | 3 | | | 4 | | | 4 | | | 6 | | | 9 | | | 6 | | | 10 | | | 13 | | |
| | $\Delta\theta$ (deg.) | 3.6 | 7.2 | 14.3 | 2.4 | 4.8 | 9.5 | 1.8 | 3.6 | 7.2 | 3.6 | 7.2 | 14.3 | 2.4 | 4.8 | 9.5 | 1.8 | 3.6 | 7.2 | 3.6 | 7.2 | 14.3 | 2.4 | 4.8 | 9.5 | 1.8 | 3.6 | 7.2 |
| 10 300 | $F$ | 11 | | | | | | | | | 11 | | | | | | | | | 11 | | | | | | | | |
| | $\delta\nu$ (GHz) | 32.2 | 16.1 | 8 | 32.2 | 16.1 | 8 | 32.2 | 16.1 | 8 | 16.1 | 8 | 4 | 16.1 | 8.0 | 4.0 | 16.1 | 8 | 4 | 10.7 | 5.4 | 2.7 | 10.7 | 5.4 | 2.7 | 10.7 | 5.4 | 2.7 |
| | $\delta\tilde{\nu}$ (cm$^{-1}$) | 1.07 | 0.54 | 0.27 | 1.07 | 0.54 | 0.27 | 1.07 | 0.54 | 0.27 | 0.54 | 0.27 | 0.13 | 0.54 | 0.27 | 0.13 | 0.54 | 0.27 | 0.13 | 0.36 | 0.18 | 0.09 | 0.36 | 0.18 | 0.09 | 0.36 | 0.18 | 0.09 |
| | $W$ (mm) | 4 | | | 6 | | | 8 | | | 8 | | | 12 | | | 16 | | | 12 | | | 18 | | | 25 | | |
| | $\Delta\theta$ (deg.) | 3.6 | 7.2 | 14.3 | 2.4 | 4.8 | 9.5 | 1.8 | 3.6 | 7.2 | 3.6 | 7.2 | 14.3 | 2.4 | 4.8 | 9.5 | 1.8 | 3.6 | 7.2 | 3.6 | 7.2 | 14.3 | 2.4 | 4.8 | 9.5 | 1.8 | 3.6 | 7.2 |
| 50 1500 | $F$ | 3 | | | | | | | | | 3 | | | | | | | | | 3 | | | | | | | | |
| | $\delta\nu$ (GHz) | 32.2 | 16.1 | 8 | 32.2 | 16.1 | 8 | 32.2 | 16.1 | 8 | 16.1 | 8 | 4 | 16.1 | 8 | 4 | 16.1 | 8 | 4 | 10.7 | 5.4 | 2.7 | 10.7 | 5.4 | 2.7 | 10.7 | 5.4 | 2.7 |
| | $\delta\tilde{\nu}$ (cm$^{-1}$) | 1.07 | 0.54 | 0.27 | 1.07 | 0.54 | 0.27 | 1.07 | 0.54 | 0.27 | 0.54 | 0.27 | 0.13 | 0.54 | 0.27 | 0.13 | 0.54 | 0.27 | 0.13 | 0.36 | 0.18 | 0.09 | 0.36 | 0.18 | 0.09 | 0.36 | 0.18 | 0.09 |
| | $W$ (mm) | 15 | | | 22 | | | 30 | | | 30 | | | 45 | | | 60 | | | 45 | | | 67 | | | 90 | | |
| | $\Delta\theta$ (deg.) | 3.6 | 7.2 | 14.3 | 2.4 | 4.8 | 9.5 | 1.8 | 3.6 | 7.2 | 3.6 | 7.2 | 14.3 | 2.4 | 4.8 | 9.5 | 1.8 | 3.6 | 7.2 | 3.6 | 7.2 | 14.3 | 2.4 | 4.8 | 9.5 | 1.8 | 3.6 | 7.2 |

**Table S3.** Lookup table for expected PC-FTS performance at $\lambda_0 = 1.5$ μm ($\nu_0 = 200$ THz; $\tilde{\nu}_0 = 6667$ cm$^{-1}$). The symbols are explained in the text.

| $\Delta\nu$ (THz) $\Delta\tilde{\nu}$ (cm$^{-1}$) | $N$ (mm$^{-1}$) | 450 | | | | | | | | | 600 | | | | | | | | | 900 | | | | | | | | |
|---|---|---|---|---|---|---|---|---|---|---|---|---|---|---|---|---|---|---|---|---|---|---|---|---|---|---|---|---|
| | $l_\text{f}$ (mm) | 100 | | | 150 | | | 200 | | | 100 | | | 150 | | | 200 | | | 100 | | | 150 | | | 200 | | |
| | $D$ (mm) | 12.5 | 25 | 50 | 12.5 | 25 | 50 | 12.5 | 25 | 50 | 12.5 | 25 | 50 | 12.5 | 25 | 50 | 12.5 | 25 | 50 | 12.5 | 25 | 50 | 12.5 | 25 | 50 | 12.5 | 25 | 50 |
| 2.5 / 83 | $F$ | 81 | | | | | | | | | 81 | | | | | | | | | 81 | | | | | | | | |
| | $\delta\nu$ (GHz) | 21.5 | 10.7 | 5.4 | 21.5 | 10.7 | 5.4 | 21.5 | 10.7 | 5.4 | 16.1 | 8 | 4 | 16.1 | 8 | 4 | 16.1 | 8 | 4 | 10.7 | 5.4 | 2.7 | 10.7 | 5.4 | 2.7 | 10.7 | 5.4 | 2.7 |
| | $\delta\tilde{\nu}$ (cm$^{-1}$) | 0.72 | 0.36 | 0.18 | 0.72 | 0.36 | 0.18 | 0.72 | 0.36 | 0.18 | 0.54 | 0.27 | 0.13 | 0.54 | 0.27 | 0.13 | 0.54 | 0.27 | 0.13 | 0.36 | 0.18 | 0.09 | 0.36 | 0.18 | 0.09 | 0.36 | 0.18 | 0.09 |
| | $W$ (mm) | 1 | | | 1 | | | 2 | | | 1 | | | 2 | | | 2 | | | 2 | | | 2 | | | 3 | | |
| | $\Delta\theta$ (deg.) | 3.6 | 7.2 | 14.3 | 2.4 | 4.8 | 9.5 | 1.8 | 3.6 | 7.2 | 3.6 | 7.2 | 14.3 | 2.4 | 4.8 | 9.5 | 1.8 | 3.6 | 7.2 | 3.6 | 7.2 | 14.3 | 2.4 | 4.8 | 9.5 | 1.8 | 3.6 | 7.2 |
| 5 / 150 | $F$ | 41 | | | | | | | | | 41 | | | | | | | | | 41 | | | | | | | | |
| | $\delta\nu$ (GHz) | 21.5 | 10.7 | 5.4 | 21.5 | 10.7 | 5.4 | 21.5 | 10.7 | 5.4 | 16.1 | 8 | 4 | 16.1 | 8 | 4 | 16.1 | 8 | 4 | 10.7 | 5.4 | 2.7 | 10.7 | 5.4 | 2.7 | 10.7 | 5.4 | 2.7 |
| | $\delta\tilde{\nu}$ (cm$^{-1}$) | 0.72 | 0.36 | 0.18 | 0.72 | 0.36 | 0.18 | 0.72 | 0.36 | 0.18 | 0.54 | 0.27 | 0.13 | 0.54 | 0.27 | 0.13 | 0.54 | 0.27 | 0.13 | 0.36 | 0.18 | 0.09 | 0.36 | 0.18 | 0.09 | 0.36 | 0.18 | 0.09 |
| | $W$ (mm) | 2 | | | 2 | | | 3 | | | 2 | | | 3 | | | 4 | | | 3 | | | 5 | | | 7 | | |
| | $\Delta\theta$ (deg.) | 3.6 | 7.2 | 14.3 | 2.4 | 4.8 | 9.5 | 1.8 | 3.6 | 7.2 | 3.6 | 7.2 | 14.3 | 2.4 | 4.8 | 9.5 | 1.8 | 3.6 | 7.2 | 3.6 | 7.2 | 14.3 | 2.4 | 4.8 | 9.5 | 1.8 | 3.6 | 7.2 |
| 10 / 300 | $F$ | 20 | | | | | | | | | 20 | | | | | | | | | 20 | | | | | | | | |
| | $\delta\nu$ (GHz) | 21.5 | 10.7 | 5.4 | 21.5 | 10.7 | 5.4 | 21.5 | 10.7 | 5.4 | 16.1 | 8 | 4 | 16.1 | 8 | 4 | 16.1 | 8 | 4 | 10.7 | 5.4 | 2.7 | 10.7 | 5.4 | 2.7 | 10.7 | 5.4 | 2.7 |
| | $\delta\tilde{\nu}$ (cm$^{-1}$) | 0.72 | 0.36 | 0.18 | 0.72 | 0.36 | 0.18 | 0.72 | 0.36 | 0.18 | 0.54 | 0.27 | 0.13 | 0.54 | 0.27 | 0.13 | 0.54 | 0.27 | 0.13 | 0.36 | 0.18 | 0.09 | 0.36 | 0.18 | 0.09 | 0.36 | 0.18 | 0.09 |
| | $W$ (mm) | 3 | | | 5 | | | 6 | | | 4 | | | 6 | | | 9 | | | 6 | | | 10 | | | 13 | | |
| | $\Delta\theta$ (deg.) | 3.6 | 7.2 | 14.3 | 2.4 | 4.8 | 9.5 | 1.8 | 3.6 | 7.2 | 3.6 | 7.2 | 14.3 | 2.4 | 4.8 | 9.5 | 1.8 | 3.6 | 7.2 | 3.6 | 7.2 | 14.3 | 2.4 | 4.8 | 9.5 | 1.8 | 3.6 | 7.2 |
| 50 / 1500 | $F$ | 5 | | | | | | | | | 5 | | | | | | | | | 5 | | | | | | | | |
| | $\delta\nu$ (GHz) | 21.5 | 10.7 | 5.4 | 21.5 | 10.7 | 5.4 | 21.5 | 10.7 | 5.4 | 16.1 | 8 | 4 | 16.1 | 8 | 4 | 16.1 | 8 | 4 | 10.7 | 5.4 | 2.7 | 10.7 | 5.4 | 2.7 | 10.7 | 5.4 | 2.7 |
| | $\delta\tilde{\nu}$ (cm$^{-1}$) | 0.72 | 0.36 | 0.18 | 0.72 | 0.36 | 0.18 | 0.72 | 0.36 | 0.18 | 0.54 | 0.27 | 0.13 | 0.54 | 0.27 | 0.13 | 0.54 | 0.27 | 0.13 | 0.36 | 0.18 | 0.09 | 0.36 | 0.18 | 0.09 | 0.36 | 0.18 | 0.09 |
| | $W$ (mm) | 13 | | | 20 | | | 27 | | | 18 | | | 27 | | | 36 | | | 27 | | | 40 | | | 54 | | |
| | $\Delta\theta$ (deg.) | 3.6 | 7.2 | 14.3 | 2.4 | 4.8 | 9.5 | 1.8 | 3.6 | 7.2 | 3.6 | 7.2 | 14.3 | 2.4 | 4.8 | 9.5 | 1.8 | 3.6 | 7.2 | 3.6 | 7.2 | 14.3 | 2.4 | 4.8 | 9.5 | 1.8 | 3.6 | 7.2 |

**Table S4.** Lookup table for expected PC-FTS performance at $\lambda_0 = 1$ μm ($\nu_0 = 300$ THz; $\tilde{\nu}_0 = 10000$ cm$^{-1}$). The symbols are explained in the text.

| $\Delta\nu$ (THz) $\Delta\tilde{\nu}$ (cm$^{-1}$) | $N$ (mm$^{-1}$) | 600 | | | | | | | | | 900 | | | | | | | | | 1200 | | | | | | | | |
|---|---|---|---|---|---|---|---|---|---|---|---|---|---|---|---|---|---|---|---|---|---|---|---|---|---|---|---|---|
| | $l_f$ (mm) | 100 | | | 150 | | | 200 | | | 100 | | | 150 | | | 200 | | | 100 | | | 150 | | | 200 | | |
| | $D$ (mm) | 12.5 | 25 | 50 | 12.5 | 25 | 50 | 12.5 | 25 | 50 | 12.5 | 25 | 50 | 12.5 | 25 | 50 | 12.5 | 25 | 50 | 12.5 | 25 | 50 | 12.5 | 25 | 50 | 12.5 | 25 | 50 |
| 2.5 / 83 | $F$ | 121 | | | | | | | | | 121 | | | | | | | | | 121 | | | | | | | | |
| | $\delta\nu$ (GHz) $\delta\tilde{\nu}$ (cm$^{-1}$) | 24.1 / 0.81 | 12.1 / 0.4 | 6 / 0.2 | 24.1 / 0.81 | 12.1 / 0.4 | 6 / 0.2 | 24.1 / 0.81 | 12.1 / 0.4 | 6 / 0.2 | 16.1 / 0.54 | 8 / 0.27 | 4 / 0.13 | 16.1 / 0.54 | 8 / 0.27 | 4 / 0.13 | 16.1 / 0.54 | 8 / 0.27 | 4 / 0.13 | 12.1 / 0.4 | 6 / 0.2 | 3 / 0.1 | 12.1 / 0.4 | 6 / 0.2 | 3 / 0.1 | 12.1 / 0.4 | 6 / 0.2 | 3 / 0.1 |
| | $W$ (mm) | 0.5 | | | 1 | | | 1 | | | 1 | | | 1 | | | 1 | | | 1 | | | 1 | | | 2 | | |
| | $\Delta\theta$ (deg.) | 3.6 | 7.2 | 14.3 | 2.4 | 4.8 | 9.5 | 1.8 | 3.6 | 7.2 | 3.6 | 7.2 | 14.3 | 2.4 | 4.8 | 9.5 | 1.8 | 3.6 | 7.2 | 3.6 | 7.2 | 14.3 | 2.4 | 4.8 | 9.5 | 1.8 | 3.6 | 7.2 |
| 5 / 150 | $F$ | 61 | | | | | | | | | 61 | | | | | | | | | 61 | | | | | | | | |
| | $\delta\nu$ (GHz) $\delta\tilde{\nu}$ (cm$^{-1}$) | 24.1 / 0.81 | 12.1 / 0.4 | 6 / 0.2 | 24.1 / 0.81 | 12.1 / 0.4 | 6 / 0.2 | 24.1 / 0.81 | 12.1 / 0.4 | 6 / 0.2 | 16.1 / 0.54 | 8 / 0.27 | 4 / 0.13 | 16.1 / 0.54 | 8 / 0.27 | 4 / 0.13 | 16.1 / 0.54 | 8 / 0.27 | 4 / 0.13 | 12.1 / 0.4 | 6 / 0.2 | 3 / 0.1 | 12.1 / 0.4 | 6 / 0.2 | 3 / 0.1 | 12.1 / 0.4 | 6 / 0.2 | 3 / 0.1 |
| | $W$ (mm) | 1 | | | 1 | | | 2 | | | 1 | | | 2 | | | 3 | | | 2 | | | 3 | | | 4 | | |
| | $\Delta\theta$ (deg.) | 3.6 | 7.2 | 14.3 | 2.4 | 4.8 | 9.5 | 1.8 | 3.6 | 7.2 | 3.6 | 7.2 | 14.3 | 2.4 | 4.8 | 9.5 | 1.8 | 3.6 | 7.2 | 3.6 | 7.2 | 14.3 | 2.4 | 4.8 | 9.5 | 1.8 | 3.6 | 7.2 |
| 10 / 300 | $F$ | 31 | | | | | | | | | 31 | | | | | | | | | 31 | | | | | | | | |
| | $\delta\nu$ (GHz) $\delta\tilde{\nu}$ (cm$^{-1}$) | 24.1 / 0.81 | 12.1 / 0.4 | 6 / 0.2 | 24.1 / 0.81 | 12.1 / 0.4 | 6 / 0.2 | 24.1 / 0.81 | 12.1 / 0.4 | 6 / 0.2 | 16.1 / 0.54 | 8 / 0.27 | 4 / 0.13 | 16.1 / 0.54 | 8 / 0.27 | 4 / 0.13 | 16.1 / 0.54 | 8 / 0.27 | 4 / 0.13 | 12.1 / 0.4 | 6 / 0.2 | 3 / 0.1 | 12.1 / 0.4 | 6 / 0.2 | 3 / 0.1 | 12.1 / 0.4 | 6 / 0.2 | 3 / 0.1 |
| | $W$ (mm) | 2 | | | 3 | | | 4 | | | 3 | | | 4 | | | 6 | | | 4 | | | 6 | | | 8 | | |
| | $\Delta\theta$ (deg.) | 3.6 | 7.2 | 14.3 | 2.4 | 4.8 | 9.5 | 1.8 | 3.6 | 7.2 | 3.6 | 7.2 | 14.3 | 2.4 | 4.8 | 9.5 | 1.8 | 3.6 | 7.2 | 3.6 | 7.2 | 14.3 | 2.4 | 4.8 | 9.5 | 1.8 | 3.6 | 7.2 |
| 50 / 1500 | $F$ | 7 | | | | | | | | | 7 | | | | | | | | | 7 | | | | | | | | |
| | $\delta\nu$ (GHz) $\delta\tilde{\nu}$ (cm$^{-1}$) | 24.1 / 0.81 | 12.1 / 0.4 | 6 / 0.2 | 24.1 / 0.81 | 12.1 / 0.4 | 6 / 0.2 | 24.1 / 0.81 | 12.1 / 0.4 | 6 / 0.2 | 16.1 / 0.54 | 8 / 0.27 | 4 / 0.13 | 16.1 / 0.54 | 8 / 0.27 | 4 / 0.13 | 16.1 / 0.54 | 8 / 0.27 | 4 / 0.13 | 12.1 / 0.4 | 6 / 0.2 | 3 / 0.1 | 12.1 / 0.4 | 6 / 0.2 | 3 / 0.1 | 12.1 / 0.4 | 6 / 0.2 | 3 / 0.1 |
| | $W$ (mm) | 9 | | | 13 | | | 17 | | | 13 | | | 19 | | | 26 | | | 17 | | | 26 | | | 34 | | |
| | $\Delta\theta$ (deg.) | 3.6 | 7.2 | 14.3 | 2.4 | 4.8 | 9.5 | 1.8 | 3.6 | 7.2 | 3.6 | 7.2 | 14.3 | 2.4 | 4.8 | 9.5 | 1.8 | 3.6 | 7.2 | 3.6 | 7.2 | 14.3 | 2.4 | 4.8 | 9.5 | 1.8 | 3.6 | 7.2 |

Table S5. Lookup table for expected PC-FTS performance at $\lambda_0 = 0.6$ μm ($\nu_0 = 500$ THz; $\tilde{\nu}_0 = 16667$ cm$^{-1}$). The symbols are explained in the text.

| $\Delta\nu$ (THz) $\Delta\tilde{\nu}$ (cm$^{-1}$) | $N$ (mm$^{-1}$) | 600 | | | | | | | | | 900 | | | | | | | | | 1200 | | | | | | | | |
|---|---|---|---|---|---|---|---|---|---|---|---|---|---|---|---|---|---|---|---|---|---|---|---|---|---|---|---|---|
| | $l_f$ (mm) | 100 | | | 150 | | | 200 | | | 100 | | | 150 | | | 200 | | | 100 | | | 150 | | | 200 | | |
| | $D$ (mm) | 12.5 | 25 | 50 | 12.5 | 25 | 50 | 12.5 | 25 | 50 | 12.5 | 25 | 50 | 12.5 | 25 | 50 | 12.5 | 25 | 50 | 12.5 | 25 | 50 | 12.5 | 25 | 50 | 12.5 | 25 | 50 |
| 2.5 / 83 | $F$ | 201 | | | | | | | | | 201 | | | | | | | | | 201 | | | | | | | | |
| | $\delta\nu$ (GHz) | 40.2 | 20.1 | 10.1 | 40.2 | 20.1 | 10.1 | 40.2 | 20.1 | 10.1 | 26.8 | 13.4 | 6.7 | 26.8 | 13.4 | 6.7 | 26.8 | 13.4 | 6.7 | 20.1 | 10.1 | 5 | 20.1 | 10.1 | 5 | 20.1 | 10.1 | 5 |
| | $\delta\tilde{\nu}$ (cm$^{-1}$) | 1.34 | 0.67 | 0.34 | 1.34 | 0.67 | 0.34 | 1.34 | 0.67 | 0.34 | 0.89 | 0.45 | 0.22 | 0.89 | 0.45 | 0.22 | 0.89 | 0.45 | 0.22 | 0.67 | 0.34 | 0.17 | 0.67 | 0.34 | 0.17 | 0.67 | 0.34 | 0.17 |
| | $W$ (mm) | 0.5 | | | 0.5 | | | 0.5 | | | 0.5 | | | 0.5 | | | 1 | | | 0.5 | | | 1 | | | 1 | | |
| | $\Delta\theta$ (deg.) | 3.6 | 7.2 | 14.3 | 2.4 | 4.8 | 9.5 | 1.8 | 3.6 | 7.2 | 3.6 | 7.2 | 14.3 | 2.4 | 4.8 | 9.5 | 1.8 | 3.6 | 7.2 | 3.6 | 7.2 | 14.3 | 2.4 | 4.8 | 9.5 | 1.8 | 3.6 | 7.2 |
| 5 / 150 | $F$ | 101 | | | | | | | | | 101 | | | | | | | | | 101 | | | | | | | | |
| | $\delta\nu$ (GHz) | 40.2 | 20.1 | 10.1 | 40.2 | 20.1 | 10.1 | 40.2 | 20.1 | 10.1 | 26.8 | 13.4 | 6.7 | 26.8 | 13.4 | 6.7 | 26.8 | 13.4 | 6.7 | 20.1 | 10.1 | 5 | 20.1 | 10.1 | 5 | 20.1 | 10.1 | 5 |
| | $\delta\tilde{\nu}$ (cm$^{-1}$) | 1.34 | 0.67 | 0.34 | 1.34 | 0.67 | 0.34 | 1.34 | 0.67 | 0.34 | 0.89 | 0.45 | 0.22 | 0.89 | 0.45 | 0.22 | 0.89 | 0.45 | 0.22 | 0.67 | 0.34 | 0.17 | 0.67 | 0.34 | 0.17 | 0.67 | 0.34 | 0.17 |
| | $W$ (mm) | 0.5 | | | 1 | | | | | | 1 | | | 1 | | | 1 | | | 1 | | | 1 | | | 1 | | |
| | $\Delta\theta$ (deg.) | 3.6 | 7.2 | 14.3 | 2.4 | 4.8 | 9.5 | 1.8 | 3.6 | 7.2 | 3.6 | 7.2 | 14.3 | 2.4 | 4.8 | 9.5 | 1.8 | 3.6 | 7.2 | 3.6 | 7.2 | 14.3 | 2.4 | 4.8 | 9.5 | 1.8 | 3.6 | 7.2 |
| 10 / 300 | $F$ | 51 | | | | | | | | | 51 | | | | | | | | | 51 | | | | | | | | |
| | $\delta\nu$ (GHz) | 40.2 | 20.1 | 10.1 | 40.2 | 20.1 | 10.1 | 40.2 | 20.1 | 10.1 | 26.8 | 13.4 | 6.7 | 26.8 | 13.4 | 6.7 | 26.8 | 13.4 | 6.7 | 20.1 | 10.1 | 5 | 20.1 | 10.1 | 5 | 20.1 | 10.1 | 5 |
| | $\delta\tilde{\nu}$ (cm$^{-1}$) | 1.34 | 0.67 | 0.34 | 1.34 | 0.67 | 0.34 | 1.34 | 0.67 | 0.34 | 0.89 | 0.45 | 0.22 | 0.89 | 0.45 | 0.22 | 0.89 | 0.45 | 0.22 | 0.67 | 0.34 | 0.17 | 0.67 | 0.34 | 0.17 | 0.67 | 0.34 | 0.17 |
| | $W$ (mm) | 1 | | | 1 | | | 1 | | | 1 | | | 2 | | | 2 | | | 1 | | | 2 | | | 3 | | |
| | $\Delta\theta$ (deg.) | 3.6 | 7.2 | 14.3 | 2.4 | 4.8 | 9.5 | 1.8 | 3.6 | 7.2 | 3.6 | 7.2 | 14.3 | 2.4 | 4.8 | 9.5 | 1.8 | 3.6 | 7.2 | 3.6 | 7.2 | 14.3 | 2.4 | 4.8 | 9.5 | 1.8 | 3.6 | 7.2 |
| 50 / 1500 | $F$ | 11 | | | | | | | | | 11 | | | | | | | | | 11 | | | | | | | | |
| | $\delta\nu$ (GHz) | 40.2 | 20.1 | 10.1 | 40.2 | 20.1 | 10.1 | 40.2 | 20.1 | 10.1 | 26.8 | 13.4 | 6.7 | 26.8 | 13.4 | 6.7 | 26.8 | 13.4 | 6.7 | 20.1 | 10.1 | 5 | 20.1 | 10.1 | 5 | 20.1 | 10.1 | 5 |
| | $\delta\tilde{\nu}$ (cm$^{-1}$) | 1.34 | 0.67 | 0.34 | 1.34 | 0.67 | 0.34 | 1.34 | 0.67 | 0.34 | 0.89 | 0.45 | 0.22 | 0.89 | 0.45 | 0.22 | 0.89 | 0.45 | 0.22 | 0.67 | 0.34 | 0.17 | 0.67 | 0.34 | 0.17 | 0.67 | 0.34 | 0.17 |
| | $W$ (mm) | 3 | | | 5 | | | 7 | | | 5 | | | 7 | | | 10 | | | 7 | | | 10 | | | 13 | | |
| | $\Delta\theta$ (deg.) | 3.6 | 7.2 | 14.3 | 2.4 | 4.8 | 9.5 | 1.8 | 3.6 | 7.2 | 3.6 | 7.2 | 14.3 | 2.4 | 4.8 | 9.5 | 1.8 | 3.6 | 7.2 | 3.6 | 7.2 | 14.3 | 2.4 | 4.8 | 9.5 | 1.8 | 3.6 | 7.2 |

## Supplementary References